\documentclass[english,aps, prb, 12pt]{revtex4-2}
\pdfoutput = 1
\usepackage[T1]{fontenc}
\usepackage[latin9]{inputenc}
\usepackage{color}
\usepackage{array}
\usepackage{float}
\usepackage{booktabs}
\usepackage{multirow}
\usepackage{amsmath}
\usepackage{amssymb}
\usepackage{stackrel}
\usepackage{graphicx}
\makeatletter

\newcommand{\lyxmathsym}[1]{\ifmmode\begingroup\def\b@ld{bold}
  \text{\ifx\math@version\b@ld\bfseries\fi#1}\endgroup\else#1\fi}

\providecommand{\tabularnewline}{\\}

\makeatother

\usepackage{babel}
\begin{document}
\title{Two-dimensional transition-metal dichalcogenides-based bilayer heterojunctions for efficient solar cells and photocatalytic applications}
\author{Khushboo Dange}
\email{khushboodange@gmail.com}

\affiliation{Department of Physics, Indian Institute of Technology Bombay, Powai,
Mumbai 400076, India}
\author{Rachana Yogi}
\email{yogirachana04@gmail.com}

\affiliation{Department of Physics, Indian Institute of Technology Bombay, Powai,
Mumbai 400076, India}
\author{Alok Shukla}
\email{shukla@iitb.ac.in}

\affiliation{Department of Physics, Indian Institute of Technology Bombay, Powai,
Mumbai 400076, India}
\begin{abstract}
In this work, we present a first-principles investigation of the optoelectronic
properties of vertically-stacked bilayer heterostructures composed
of 2D transition-metal dichalcogenides (TMDs). The calculations are
performed using the density-functional theory (DFT) as well as many-body
perturbation theory within G$_{0}$W$_{0}$-BSE methodology. Our aim
is to propose these TMD heterostructures for potential applications
in solar cells. The TMD monolayers comprising the heterojunctions
considered in this research are MoS\textsubscript{2}, WS\textsubscript{2},
MoSe\textsubscript{2}, and WSe\textsubscript{2} due to their favorable
band gaps, high carrier mobility, robust absorption in the visible
region, and excellent stability. These four TMD monolayers provide
the basis for a total of six potential heterostructures. Consequently,
we have examined the structural, electronic, and optical properties
of six heterostructures (WS\textsubscript{2}/MoS\textsubscript{2},
MoSe\textsubscript{2}/MoS\textsubscript{2}, MoSe\textsubscript{2}/WS\textsubscript{2},
WSe\textsubscript{2}/MoS\textsubscript{2}, WSe\textsubscript{2}/MoSe\textsubscript{2},
and WSe\textsubscript{2}/WS\textsubscript{2}). At the DFT level,
all the six considered TMD heterostructures meet the essential criterion
of type II band alignment, a critical factor in extending carrier
lifetime. However, according to G$_{0}$W$_{0}$ results, MoSe\textsubscript{2}/WS\textsubscript{2}
does not exhibit the type II band alignment, instead it shows type
I band alignment. The significantly large quasi particle gaps obtained
from G$_{0}$W$_{0}$ approximation suggest the presence of strong
electron-correlation effects. The chosen heterostructures exhibit
superior optoelectronic properties compared to their respective isolated
monolayers. The quite significant values of the intrinsic electric
fields which arise due to the asymmetric geometry of the heterostructures
are obtained. Additionally, the obtained small and nearly equal electron
and hole effective masses indicate high mobility and efficient charge
carrier separation, resulting in low recombination losses. The quality
of these heterojunction solar cells is estimated by computing their
power conversion efficiencies (PCE). The PCEs are calculated at both
the HSE06 and G$_{0}$W$_{0}$ levels, and the maximum PCE predicted
by HSE06 calculations on our designed solar cells can reach up to
19.25\% for the WSe\textsubscript{2}/WS\textsubscript{2} heterojunction.
In addition, all the six TMD heterostructures are examined for their
potential applications in photocatalysis for hydrogen evolution reaction,
and the three of them, namely, WS\textsubscript{2}/MoS\textsubscript{2},
MoSe\textsubscript{2}/MoS\textsubscript{2}, and WSe\textsubscript{2}/MoS\textsubscript{2}
heterostructures qualify for the same.
\end{abstract}
\maketitle

\section{Introduction}

As traditional fossil fuel-based sources are proving to be unsustainable
and detrimental to the environment, non-conventional energy resources
are emerging as viable alternatives. Solar energy is one such non-conventional
energy resource, which provides a clean, renewable, and abundant alternative
\citep{Gong2019}. The electricity can be generated by harnessing
the power of the sun without emitting harmful greenhouse gases, thereby
reducing our carbon footprint and mitigating climate change. With
advancing technology and decreasing costs, the use of solar energy
can drive economic growth and empower communities to become self-sufficient
in meeting their energy needs. However, designing devices for harvesting
solar power on a large scale in a commercially viable manner continues
to be a big challenge for scientists \citep{ahmad2021}. 

2D materials have opened up exciting possibilities for enhancing efficiency
of harvesting solar energy. These ultrathin materials such as graphene
\citep{IQBAL2018}, transition metal dichalcogenides (TMDs) \citep{Jariwala2014,ganesan2016},
and perovskites \citep{Zhang2022}, possess desirable optoelectronic
properties, making them highly suitable for solar cell applications.
Their large surface area and excellent electrical conductivity \citep{Wu2009,Manzeli2017}
allow for efficient charge separation, resulting in improved performance.
Moreover, 2D materials can be engineered to absorb light across a
broad spectrum, enabling them to capture a wider range of solar radiations.
With the advancement of novel 2D materials, their heterojunctions
held together by van der Waals (vdW) interactions have been demonstrated
both theoretically and practically, not just to probe the fundamental
physics, but also for their potential device applications \citep{Novoselov2016,Geim2013}.
The weak vdW interlayer forces not only retain several desirable intrinsic
properties of the stacked structures, but also result in new ones
such as the increased optical absorption range \citep{Wu2012}, charge
transfer \citep{Du2012}, band alignment \citep{Zhang2016}, etc.
Heterojunctions with type II band alignment play a significant role
in increasing solar energy harvesting efficiency. The presence of
a type-II band alignment in a heterostructure comprising two semiconductors,
with one serving as a donor and the other one as an acceptor, results
in the segregation of photoexcited electrons and holes at the interface.
This results in an indirect exciton formation, which thereby reduces
the carrier recombination rate. 

Among the 2D material family, TMDs are emerging as promising candidates
for applications in catalysis, energy storage, gas sensing, field-effect
transistors, logic circuits, and optoelectronics particularly due
to their semiconducting characteristics, strong chemical stability,
and flexibility \citep{CHOI2017116,Ahmed2017,Wu2023,Qian2022,Wang2012}.
Bulk TMDs formed by transition metals (M = Mo, W) and chalcogens (X
= S, Se) exist in two structural phases i.e., trigonal prismatic (2H)
and octahedral (1T) phases depending on the different stacking order
of the three atomic planes (X-M-X) \citep{Manzeli2017}. Both the
2H and 1T phases are thermodynamically stable. The synthesis of monolayer
TMDs has already been reported in various studies \citep{faiha2023,McCreary2016,Wang2014,Huang2014}.
These TMDs undergo indirect to direct band gap transition on reducing
the thickness from bulk to monolayer \citep{mos2exp,Splendiani2010}.
The direct band gap in the visible region of the spectrum, high carrier
mobility, exotic optical properties, and high stability in ambient
conditions make these TMD monolayers potential candidates to be used
in heterojunction solar cells \citep{Mak2016}. Several heterojunction
solar cell devices composed of other 2D materials with TMDs have been
proposed theoretically or realized experimentally, such as graphene/MoS\textsubscript{2}
\citep{Bernardi2013}, phosphorene/MoS\textsubscript{2} \citep{Dai2014},
MoSe\textsubscript{2}/$\psi$-phosphorene \citep{Wangpsi2017}, and
tellurene/MX\textsubscript{2} (M=Mo, W; X=S, Se, Te) \citep{WuKai2019}.
Heterostructures in which both the considered 2D materials are TMDs
have also been investigated both experimentally and theoretically
\citep{kim2020,Chen2017,Amin2015,Ceballos2014}. Kim \emph{et al.}
\citep{kim2020} investigated the sensing properties of the WSe\textsubscript{2}/WS\textsubscript{2}
and MoS\textsubscript{2}/WSe\textsubscript{2} heterojunctions for
NO\textsubscript{2} and NH\textsubscript{3} gases. Their study confirmed
that 2D TMD heterostructures possess potential for use in gas sensor
applications without external biasing. Chen \emph{et al.} \citep{Chen2017}
synthesized some 2D TMD heterostructures on Si/SiO\textsubscript{2}
substrates and explored their Raman and photoluminescence spectra.
The successful synthesis of various 2D TMD heterostructures using
controllable interfaces provides exciting opportunities to design
a variety of novel optoelectronic devices. Amin \emph{et al.} \citep{Amin2015}
studied the electronic and optical properties of out-of-plane and
in-plane heterostructures of 2D TMDs and predicted their potential
for photocatalytic applications. 

The present study employs first-principles density-functional theory
calculations augmented by the inclusion of the electron-correlation
effects within a G$_{0}$W$_{0}$-- BSE approach to study the TMD
vertical heterostructures for their possible applications in solar
cells. By considering all possible combinations, in total six heterostructures
are formed using the 1H phases of four TMDs monolayers (MX\textsubscript{2},
where M = Mo, W; X = S, Se). The electronic and optical properties
of all the heterostructures are investigated in detail, along with
the calculation of their power conversion efficiencies. All the six
considered heterostructures exhibit type II band alignment at the
HSE06 level, whereas, at the G$_{0}$W$_{0}$ level, MoSe\textsubscript{2}/WS\textsubscript{2}
heterostructure displays a type I band alignment, while the rest still
have a type II band alignment. After thoroughly examining the properties
relevant for photovoltaics, we delve into a study of the photocatalytic
performance of all the six considered TMD heterostructures. To the
best of our knowledge, a comparative study of this type both at the
HSE06, as well as G$_{0}$W$_{0}$-- BSE levels, has not been reported
so far.

The remainder of this paper is organized in the following manner.
The next section contains the details of the employed computational
tools in brief, followed by a systematic presentation and discussion
of our results in Sec. \ref{sec:results-and-discussion}. Finally,
we conclude by summarizing our findings in Sec. \ref{sec:Conclusion}. 

\section{Computational Methods}

Our calculations were performed using the plane-wave based Vienna
ab initio simulation package (VASP) \citep{kresse1996,Georg1996}
which employs a first-principles based density functional theory (DFT)
\citep{hohenberg1964,kohn1965}. Further, Perdew--Burke--Ernzerhof
(PBE) \citep{perdew1996} exchange correlation functional of the generalized
gradient approximation (GGA) is used in all the calculations. The
electronic structure calculations are also performed using the Heyd-Scuseria-Ernzerhof
(HSE06) \citep{heyd2003hybrid} hybrid functional to overcome the
band gap under estimation by the GGA functionals. Since both these
functionals do not correctly describe the vdW interactions, we adopted
the non-local optB88-vdW \citep{Klime2010} dispersion corrections
to account for them in the considered heterostructures. The projected
augmented wave (PAW) \citep{kresse1999,blochl1994} method is used,
which includes pseudopotentials coupled with the plane wave basis
set. In these calculations, we used a relatively large basis set with
the kinetic energy cutoff of 600 eV. Our calculations were performed
on the primitive cells so that the monolayer calculations involved
three atoms, while the heterostructure calculations were performed
using six atoms. The vacuum of 20 $\lyxmathsym{\AA}$ and 25 $\lyxmathsym{\AA}$
are created along the $z$-direction in the case of monolayers and
heterostructures, respectively, to avoid the interaction with their
periodic copies. The geometries are optimized until the Hellman-Feynman
force acting on each atom is less than 0.01 $\text{eV/\AA}$,
while the convergence criterion for the self-consistent-field (SCF)
iterations was $10^{-5}$ eV on the total energy of the system.  The
Brillouin zone integration is performed using the Monkhorst-Pack k-points
grid \citep{monkhorst1976}, and meshes of the sizes $9\times9\times1$
and $9\times9\times2$ are employed for the geometry optimization
of the TMD monolayers and heterostructures, respectively. Next, the
refined k-mesh of $12\times12\times1$ is used for further calculations
on the optimized monolayers, but in case of heterojunctions, the k-mesh
is not refined due to the high computational cost. The effect of spin-orbit
coupling (SOC) is also taken into consideration in both the GGA-PBE
and HSE06 calculations on the optimized structures as it is significant
for both the Mo and W atoms. Despite the fact that the TMD monolayers
under consideration are non-magnetic in nature, effect of spin-polarization
is taken into account during the optimization of the heterostructures,
which verifies their non-magnetic nature. Following this, all the
calculations are performed under a non-spin-polarized condition.

To calculate the quasiparticle (QP) energies, we employed the many-body
perturbation theory using the GW approximation \citep{Shiskin2006}.
Specifically, we performed single-shot G$_{0}$W$_{0}$ calculations
based on the Bloch functions obtained from the GGA-PBE calculations,
and a kinetic-energy cutoff of 400 eV for the response function was
employed. We included 160 unoccupied bands for the G$_{0}$W$_{0}$
calculations which are further increased to 200 to check the convergence
of our results. Notably, the QP band gap values obtained with 200
unoccupied bands were nearly identical to those with 160 bands. As
a result, we proceeded with 160 unoccupied bands for further calculations.
To obtain the G$_{0}$W$_{0}$ band structures, we used the Wannier
interpolation approach, as implemented in Wannier90 program \citep{Arash2008}.
For the Wannier interpolation, the initial projections were chosen
as the\textit{ d} orbitals of transition metals (Mo, W) and \textit{p}
orbitals of chalcogens (S, Se). The band alignment at the G$_{0}$W$_{0}$
level is obtained using the band-gap center as a reference,
as it has been shown to be relatively insensitive to the choice of
functional \citep{Houlong2013}. According to the method proposed
by Toroker \textit{et al.} \citep{Toroker2011}, QP band edges were
determined as 

\begin{equation}
E_{CBM/VBM}=E_{BGC}\pm\frac{1}{2}E_{g}^{QP},\label{eq:BGC}
\end{equation}

where $E_{g}^{QP}$ denotes the G$_{0}$W$_{0}$ band gap, $E_{BGC}$ is the band-gap center, and $E_{CBM}$
and $E_{VBM}$, respectively, denote the positions of the valence
band maximum (VBM) and conduction band minimum (CBM) with respect
to the vacuum, respectively. 

For the calculation of the optical absorption spectra, we solve the
Bethe-Salpeter equation (BSE) \citep{Albrecht1998,Rohlfing2000} 

\begin{equation}
(E_{c\boldsymbol{k}}-E_{v\boldsymbol{k}})A_{vc\boldsymbol{k}}^{S}+\underset{\boldsymbol{k'v'c'}}{\sum}\left\langle vc\boldsymbol{k}\left|K_{eh}\right|v'c'\boldsymbol{k'}\right\rangle A_{v'c'\boldsymbol{k'}}^{S}=\Omega^{S}A_{vc\boldsymbol{k}}^{S}\label{eq:BSE-eq}
\end{equation}

where $E_{v\boldsymbol{k}}$ and $E_{c\boldsymbol{k}}$ are the QP
energies of the valence and conduction band states, respectively,
$A_{vc\boldsymbol{k}}^{S}$ is the electron-hole amplitude matrix
corresponding to transition S, $\Omega^{S}$ represents the eigenenergies
corresponding to the transition, while $K_{eh}$ is the interaction
kernel. Eq. \ref{eq:BSE-eq} are solved within the Tamm-Dancoff approximation\citep{Taylor1954},
using the single-particle states obtained from the G$_{0}$W$_{0}$
calculations. We found that the four topmost valence bands and four
lowest conduction bands as the basis for the excitonic states in the
Bethe-Salpeter kernel ( $K_{eh}$) are sufficient for the relatively
low-energy optical excitations that we are interested in.

\section{\protect\label{sec:results-and-discussion}results and discussion}

\subsection{Electronic structure of TMD Monolayers}

\begin{figure}[H]
\begin{centering}
\includegraphics[scale=0.3]{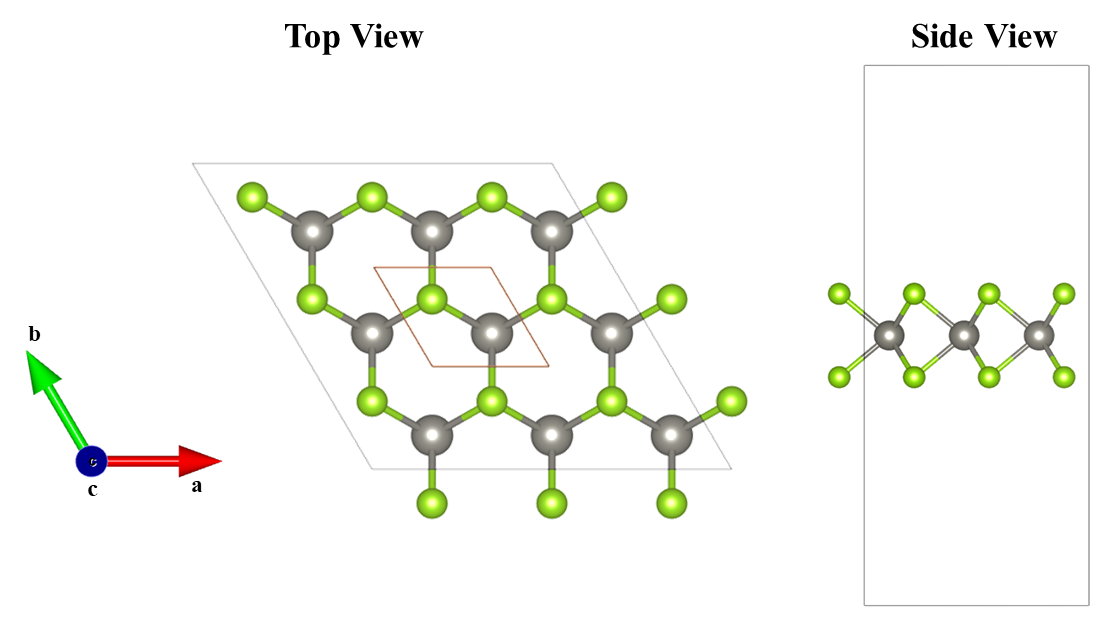}
\par\end{centering}
\caption{\protect\label{fig:mono_struc}Structure of the considered TMD monolayers.
Here, grey and green spheres represent the M (=Mo or W) and X (=S
or Se) atoms, respectively.}
\end{figure}

We first considered TMD monolayers (MoS\textsubscript{2}, MoSe\textsubscript{2},
WS\textsubscript{2}, and WSe\textsubscript{2}) in their 1H phase,
and calculated their electronic properties using the optimized structures.
A monolayer of these TMDs consists of three atomic planes, out of
which one layer of metal atoms (M) is sandwiched between two layers
of chalcogens (X). These monolayers have \textit{P6m2} space group
with the D$_{3h}$ point group symmetry, and their optimized geometries
are shown in Fig. \ref{fig:mono_struc}, with the optimized lattice
constants reported in Table \ref{tab:Electronic=000020gap_monolayers}.
The obtained lattice constant for each of the monolayers is consistent
with the values reported in literature \citep{WuKai2019,Cslattice,Annalen2014}.
The electronic band structures are obtained using both the PBE and
HSE06 functionals, and the corresponding band gap ($E_{g}$) values
are presented in Table \ref{tab:Electronic=000020gap_monolayers}.
The calculated HSE06 band structures of the considered monolayers
are presented in Figs. S1(a)-(d) of the Supplemental Material (SM). For each of the four monolayers, a direct band
gap is obtained at K-point using both the functionals. The band gap
values and their direct nature are consistent with the previously
reported results \citep{mos2dft,WuKai2019}. The obtained upward splittings
in VBM due to SOC are 0.15 eV, 0.43 eV, 0.19 eV, and 0.47 eV for the
MoS\textsubscript{2}, WS\textsubscript{2}, MoSe\textsubscript{2},
and WSe\textsubscript{2} monolayers, respectively, in agreement with
the literature \citep{Zhang2017}. The $E_{g}$ values obtained using
the HSE06 functional are close to the experimentally reported values
mentioned in Table \ref{tab:Electronic=000020gap_monolayers}. The
direct band gap is an important factor for better optoelectronic properties.
For highly efficient heterojunction solar cells, both donor and acceptor
semiconductors should possess a desirable band gap, i.e., in the range
of 1.10 eV--1.74 eV \citep{shockley1961}, so as to guarantee a wide
sunlight absorption range. Thus, based on the band gap criterion,
MoSe\textsubscript{2} and WSe\textsubscript{2} are relatively better
candidates for heterojunction solar cells. Further, the considered
monolayers can achieve high carrier mobility (values are presented
in Table S1 of the SM), resulting in efficient
carrier transport \citep{li2019,Zhang2014}. 

\begin{table}[H]
\begin{centering}
\caption{\protect\label{tab:Electronic=000020gap_monolayers}Calculated lattice
constants and the electronic band gaps for the considered TMD monolayers.}
\par\end{centering}
\centering{}%
\begin{tabular}{cccccccc}
\toprule 
\multirow{2}{*}{S.No.} & \multirow{2}{*}{Monolayers} & Lattice Constant & \multicolumn{5}{c}{Band Gap (eV)}\tabularnewline
\cmidrule{4-8}
 &  & (\AA) & \multicolumn{1}{c}{PBE} & \multicolumn{1}{c}{HSE06} & G$_{0}$W$_{0}$ & G$_{0}$W$_{0}$+ SOC & Experimental\tabularnewline
\midrule
\midrule 
1 & MoS\textsubscript{2} & 3.180 & 1.59 \citep{mos2dft} & 2.02\citep{mos2dft} & 2.50 & 2.42 & 1.90 \citep{mos2exp,mos2exp2}\tabularnewline
2 & WS\textsubscript{2} & 3.190 & 1.51 \citep{mos2dft} & 1.96\citep{mos2dft} & 2.76 & 2.50 & 1.96 \citep{ws2exp}, 2.0 \citep{Zhao2013}\tabularnewline
3 & MoSe\textsubscript{2} & 3.320 & 1.32 \citep{WuKai2019} & 1.71\citep{WuKai2019} & 2.17 & 2.06 & 1.58 \citep{mose2exp}\tabularnewline
4 & WSe\textsubscript{2} & 3.319 & 1.24 \citep{WuKai2019} & 1.65\citep{WuKai2019} & 2.70 & 2.40 & 1.96 \citep{wse2exp}\tabularnewline
\bottomrule
\end{tabular}
\end{table}

The G$_{0}$W$_{0}$ band structures are also calculated for the considered
TMD monolayers. We first calculated the Wannier-interpolated PBE band
structures (without SOC) and compared them to the DFT band structures,
as shown in Fig. S2 of the SM for the MoS\textsubscript{2}
monolayer. After confirming that the Wannierized band structures closely
matched the DFT results, we proceeded to calculate the G$_{0}$W$_{0}$
band structures using the same set of Wannier functions. The resultant
G$_{0}$W$_{0}$ band structures of all the four monolayers are presented
in Figs. S1(e)-(h) of the SM. The corresponding
QP band gaps ($E_{g}^{QP}$) obtained from the G$_{0}$W$_{0}$ approximation
are listed in Table \ref{tab:Electronic=000020gap_monolayers}. These
band gaps, significantly larger than those obtained from HSE06 calculations,
suggest the presence of strong electron-correlation effects, and are
consistent with the previously reported values \citep{Molina2013,Zhang2017}.
For example, our $E_{g}^{QP}$ values for MoS\textsubscript{2} (2.50
eV) and MoSe\textsubscript{2} (2.17 eV) monolayers are in excellent
agreement with the values reported by Torun \textit{et al.} (2.54
eV and 2.19 eV, respectively) \citep{Torun2018}. Further, G$_{0}$W$_{0}$
approximation does not alter the nature of the band gaps, as a direct
band gap (at K) persists for all the four monolayers. Since the effect
of SOC is not included in the G$_{0}$W$_{0}$ gaps, we obtained SOC
corrected G$_{0}$W$_{0}$ gaps by rigidly shifting the CBM and VBM
by the same amount as the corresponding shifts observed in the PBE
band structures computed including the SOC, and the results are reported
in Table \ref{tab:Electronic=000020gap_monolayers}. Notably, our
SOC-corrected $E_{g}^{QP}$ for the MoS\textsubscript{2} monolayer
(2.42 eV) closely matches the gap reported by Alejandro \textit{et
al.} \citep{Molina2013} (2.41 eV) who included SOC in the SCF calculations,
thus validating our approach of applying a rigid shift to account
for SOC in the G$_{0}$W$_{0}$ gaps. Given that PBE underestimates
the band gap, we continued with the HSE06 and G$_{0}$W$_{0}$ results
to assess the band alignment between the monolayers in the heterostructures
under consideration, as well as to study their potential for solar
cell applications.

\subsection{Optical spectra of TMD monolayers}

The BSE calculations are performed to obtain the optical response
of the considered monolayers. Both the imaginary and real parts of
dielectric function ($\epsilon(\omega$)), defined as $\epsilon(\omega)=\epsilon_{Re}(\omega)+i\epsilon_{Im}(\omega)$
are obtained by performing the BSE simulations. The imaginary part
of dielectric function ($\epsilon_{Im}(\omega$)) describes the optical
absorption properties and is presented in Fig. S3 of the SM for all the monolayers. It is observed that the first peak occurs
at an energy lower than the QP gap obtained using the G$_{0}$W$_{0}$
approximation. This suggests that the optical gap, $E_{g}^{op}$,
is smaller than the electronic gap due to excitonic effects caused
by electron-hole attraction. The $E_{g}^{op}$ values obtained from
the BSE spectrum (Fig. S3 of the SM) and the corresponding
exciton binding energies, $E_{ex}$, calculated as $E_{ex}=E_{g}^{QP}-E_{g}^{op}$
are tabulated below (see Table \ref{tab:monolayer-optical-gap}).
Further, the SOC corrected optical gaps, listed in the same table
(see Table \ref{tab:monolayer-optical-gap}) are obtained by subtracting
$(E_{g})_{G_{0}W_{0}}-(E_{g})_{G_{0}W_{0}+SOC}$ from the original
$E_{g}^{op}$ values, which did not account for SOC. The SOC corrected
$E_{g}^{op}$ values thus obtained show excellent agreement with the
experimental gaps (see Table \ref{tab:Electronic=000020gap_monolayers}),
further validating our approach of accounting for SOC through a rigid
shift. Moreover, the obtained $E_{ex}\sim0.5$ eV for all the four
TMD monolayers agree well with both the theoretical \citep{Berkelbach}
and experimental studies \citep{Klots2014,Chernikov2014}. The larger
$E_{ex}$ values observed for the W based TMD monolayers are consistent
with the stronger electron correlation effects arising from the contribution
of 4\textit{f} and 5\textit{d} electrons, as compared to just 4\textit{d}
electrons in the Mo based TMD monolayers.

\begin{table}[H]
\caption{\protect\label{tab:monolayer-optical-gap}Optical gaps, $E_{g}^{op}$
obtained from the BSE calculations and the corresponding exciton binding
energies, $E_{ex}$ for the considered TMD monolayers.}

\begin{centering}
\begin{tabular}{ccccc}
\toprule 
\multirow{2}{*}{S.No.} & \multirow{2}{*}{Monolayers} & \multicolumn{2}{c}{$E_{g}^{op}$(eV)} & $E_{ex}$(eV)\tabularnewline
\cmidrule{3-5}
 &  & BSE & BSE+SOC & \tabularnewline
\midrule
\midrule 
1 & MoS\textsubscript{2} & 2.04 & 1.96 & 0.46\tabularnewline
2 & WS\textsubscript{2} & 2.27 & 2.01 & 0.49\tabularnewline
3 & MoSe\textsubscript{2} & 1.71 & 1.60 & 0.46\tabularnewline
4 & WSe\textsubscript{2} & 2.17 & 1.87 & 0.53\tabularnewline
\bottomrule
\end{tabular}
\par\end{centering}
\end{table}

\subsection{Selection Criteria for heterostructures formation }

Total six combinations of two different semiconductors are possible
using the four considered TMD monolayers. When assessing the potential
of two materials for heterostructure formation, their lattice constants
play a crucial role. The lattice mismatch ($\nu$) between the two
semiconductors can be calculated using the following formula: 

\begin{equation}
\nu=\frac{2\left|a_{1}-a_{2}\right|}{a_{1}+a_{2}}\times100\%,\label{eq:lattice_mismatch}
\end{equation}

where $a_{1}$ and $a_{2}$ are the lattice constants of the semiconductors
involved in the formation of heterostructure. Using Eq. \ref{eq:lattice_mismatch}
and the obtained lattice constants for the monolayers (Table \ref{tab:Electronic=000020gap_monolayers}),
we have calculated $\nu$ for the six possible heterostructures (see
Table \ref{tab:lattice-mismatch}). Relatively small values of $\nu$
(< 5\%) obtained for all the six combinations, combined with the fact
that all the monolayers have the same geometric symmetries suggests
that it is possible to make vertically-stacked bilayer heterojunctions
from them. As a result of lattice mismatch, the two layers get strained
during optimization.

\begin{table}[H]
\caption{\protect\label{tab:lattice-mismatch}Calculated lattice mismatch ($\nu$)
for the TMD heterostructures under consideration.}

\centering{}%
\begin{tabular}{c|cccccc}
\hline 
Heterostructures & WS\textsubscript{2}/MoS\textsubscript{2} & MoSe\textsubscript{2}/MoS\textsubscript{2} & MoSe\textsubscript{2}/WS\textsubscript{2} & WSe\textsubscript{2}/MoS\textsubscript{2} & WSe\textsubscript{2}/MoSe\textsubscript{2} & WSe\textsubscript{2}/WS\textsubscript{2}\tabularnewline
\hline 
\hline 
$\nu$ (\%) & 0.30 & 4.30 & 3.99 & 4.27 & 0.03 & 3.96\tabularnewline
\hline 
\end{tabular}
\end{table}

\begin{figure}[H]
\begin{centering}
\includegraphics[scale=0.3]{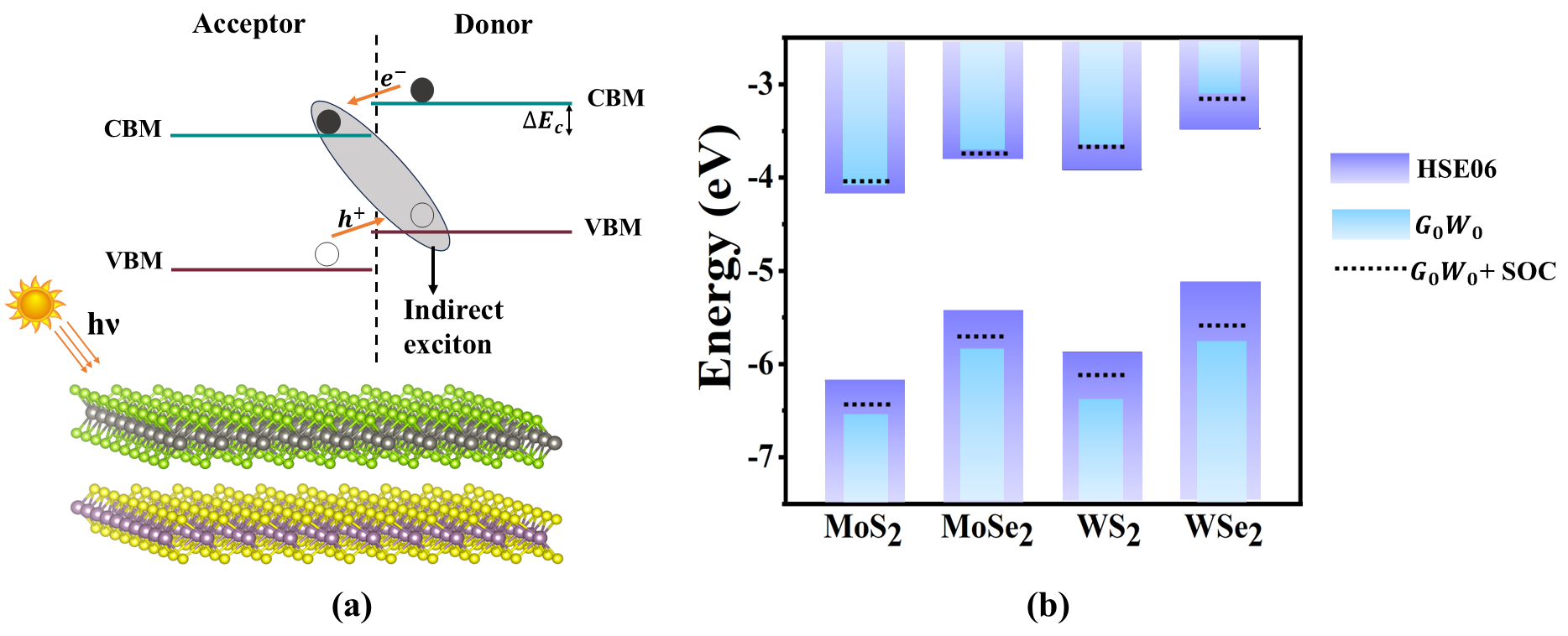}
\par\end{centering}
\caption{\protect\label{fig:type2_align}(a) Schematics of the type II band
alignment and (b) Position of the VBM and CBM energy levels of the
considered TMD monolayers with respect to vacuum under different approaches.}
\end{figure}

Another important factor is that the type II band alignment, shown
in Fig. \ref{fig:type2_align}(a), between the two chosen semiconductors
for a heterostructure is crucial for its application in solar cells.
This is because in type II band alignment, the photoexcited electrons
generated in a semiconductor (donor) are confined to one side (lowest
conduction band of the acceptor) of the heterostructure, and holes
are confined to the other side (highest occupied valence band of the
donor). This spatial separation of electrons and holes results in
an indirect exciton formation and thus reduces carrier recombination
losses, leading to longer carrier lifetimes and improved overall efficiency.
To determine the type of band alignment, band edge positions, $E_{CBM}$
and $E_{VBM}$ of the considered TMD monolayers are calculated using
the results of both the HSE06 functional and G$_{0}$W$_{0}$ approximation.
In case of G$_{0}$W$_{0}$ calculations, the band edge positions
are computed using Eq. \ref{eq:BGC}, and the calculated $E_{CBM}$
and $E_{VBM}$ at the HSE06 and G$_{0}$W$_{0}$(without and with
SOC) levels are summarized in Table \ref{tab:Band-edge-positions,},
as well as shown in an energy level diagram in Fig. \ref{fig:type2_align}(b).
It is found that the five heterostructures, MoSe\textsubscript{2}/MoS\textsubscript{2},
WS\textsubscript{2}/MoS\textsubscript{2}, WSe\textsubscript{2}/MoS\textsubscript{2},
WSe\textsubscript{2}/MoSe\textsubscript{2}, and WS\textsubscript{2}/WSe\textsubscript{2},
exhibit type II band alignment according to both the HSE06 calculations
and G$_{0}$W$_{0}$ approximation, in agreement with the literature
\citep{Zhang2017}. However, for MoSe\textsubscript{2}/WS\textsubscript{2},
HSE06 calculations predicts type II alignment, whereas, G$_{0}$W$_{0}$
(both with and without SOC) indicates type I alignment. This discrepancy
in band alignment between MoSe\textsubscript{2} and WS\textsubscript{2}
monolayers has also been reported in previous theoretical and experimental
studies \citep{Zhang2017,Ma2021,Kistner2024}. The convention A/B
is used throughout the paper, to represent the considered heterostructures,
in which A and B denote the semiconductors acting as donor and acceptor,
respectively. 

\begin{table}[H]

\begin{centering}
\caption{\textcolor{blue}{\protect\label{tab:Band-edge-positions,}}Band edge
positions, $E_{VBM}$ and $E_{CBM}$ of the considered TMD monolayers
with respect to vacuum, computed at the HSE06, G$_{0}$W$_{0}$, and
G$_{0}$W$_{0}$+ SOC level. }
\begin{tabular}{ccccccc}
\toprule 
\multirow{2}{*}{Monolayers} & \multicolumn{3}{c}{$E_{VBM}$(eV)} & \multicolumn{3}{c}{$E_{CBM}$(eV)}\tabularnewline
\cmidrule{2-7}
 & HSE06 & G$_{0}$W$_{0}$ & G$_{0}$W$_{0}$+ SOC & \ \ \ HSE06\  & G$_{0}$W$_{0}$ & G$_{0}$W$_{0}$+ SOC\tabularnewline
\midrule
\midrule 
MoS\textsubscript{2} & -6.181 & -6.559 & -6.481 & \ \ \ -4.159\  & -4.060 & -4.059\tabularnewline
WS\textsubscript{2} & -5.877 & -6.391 & -6.166 & \ \ \ -3.914\  & -3.627 & -3.665\tabularnewline
MoSe\textsubscript{2} & -5.448 & -5.846 & -5.745 & \ \ \ -3.739\  & -3.678 & -3.681\tabularnewline
WSe\textsubscript{2} & -5.127 & -5.801 & -5.544 & \ \ \ -3.475\  & -3.097 & -3.140\tabularnewline
\bottomrule
\end{tabular}
\par\end{centering}
\end{table}

\subsection{Calculated properties of TMD heterostructures}

\subsubsection{Structural Properties}

\begin{figure}[H]
\begin{centering}
\includegraphics[scale=0.35]{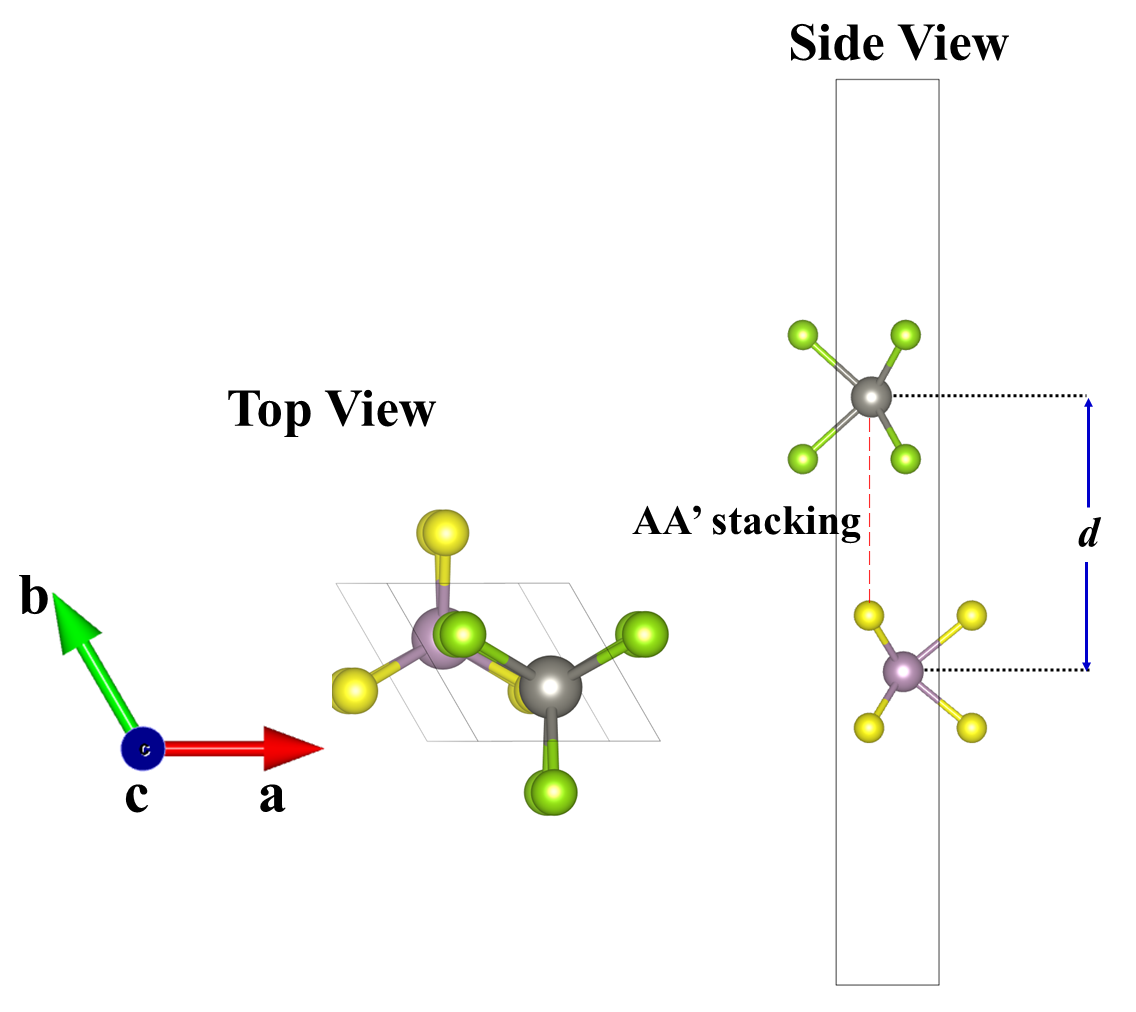}
\par\end{centering}
\caption{\protect\label{fig:heterostructunit}The vertical heterostructures
of TMDs, formed using AA' stacking. Here, purple and grey spheres
symbolize the metal atoms M (= Mo or W), while yellow and green spheres
stand for the chalcogen atoms X (=S or Se), of the two different layers.}
\end{figure}

The vertical heterostructures formed using TMD monolayers are shown
in Fig. \ref{fig:heterostructunit}. To form these heterostructures,
the AA' type of stacking is considered, in which two layers are aligned
anitparallel to each other and an X atom of the lower layer lies below
a M atom of the top layer. The lattice constants (a) of the optimized
heterostructures are presented in Table \ref{tab:struct_para}, and
on comparing these lattice constants with those of the constituent
monolayers, we found that the monolayers get strained to attain the
ground state. For instance, from the lattice constant of MoSe\textsubscript{2}/MoS\textsubscript{2}
(3.18 \AA), we infer that MoSe\textsubscript{2} layer gets compressed,
while the MoS\textsubscript{2} layer remains unstrained.
The optimized interlayer distance $d$, i.e., the perpendicular distance
between the M-atom layers of the two monolayers, is obtained in the
range of 6.0 {\AA} -- 6.3 {\AA} for all the considered heterostructures (see
Table \ref{tab:struct_para} for exact values), close to the reported
values \citep{komsa2013,Torun2018}. To determine the stability of
these heterostructures, their formation energies ($E_{f}$) are calculated
using the following formula

\begin{equation}
E_{f}=E_{hetero}-(E_{layer1}+E_{layer2})\label{eq:formation}
\end{equation}

where $E_{hetero}$ represents the total energy of the heterostructure,
while $E_{layer1}$ and $E_{layer2}$ denote the energies of the two
different TMD monolayers constituting the heterojunction. The calculated
$E_{f}$ values are presented in Table \ref{tab:struct_para}. The
negative $E_{f}$ values are obtained for all the six heterostructures
in the range -0.31 eV to -0.39 eV, confirming their stability. Further,
the weak vdW interactions between the two constituent layers of the
heterostructures are responsible for the small $E_{f}$ values.

\begin{table}[H]
\caption{\protect\label{tab:struct_para}Calculated interlayer distances ($d$),
lattice constants (a), and formation energies ($E_{f}$) for the TMD
heterostructures under consideration.}

\centering{}%
\begin{tabular}{ccccc}
\toprule 
\multirow{2}{*}{S.No.} & \multirow{2}{*}{Heterostructures} & \multirow{2}{*}{d (\AA)} & \multirow{2}{*}{a (\AA)} & \multirow{2}{*}{$E_{f}$ (eV)}\tabularnewline
 &  &  &  & \tabularnewline
\midrule
\midrule 
1 & WS\textsubscript{2}/MoS\textsubscript{2} & 6.00 & 3.12 & -0.36\tabularnewline
2 & MoSe\textsubscript{2}/MoS\textsubscript{2} & 6.11 & 3.18 & -0.32\tabularnewline
3 & MoSe\textsubscript{2}/WS\textsubscript{2} & 6.14 & 3.18 & -0.31\tabularnewline
4 & WSe\textsubscript{2}/MoS\textsubscript{2} & 6.13 & 3.18 & -0.31\tabularnewline
5 & WSe\textsubscript{2}/MoSe\textsubscript{2} & 6.30 & 3.24 & -0.39\tabularnewline
6 & WSe\textsubscript{2}/WS\textsubscript{2} & 6.14 & 3.18 & -0.31\tabularnewline
\bottomrule
\end{tabular}
\end{table}

\begin{figure}[H]
\begin{centering}
\includegraphics[scale=0.35]{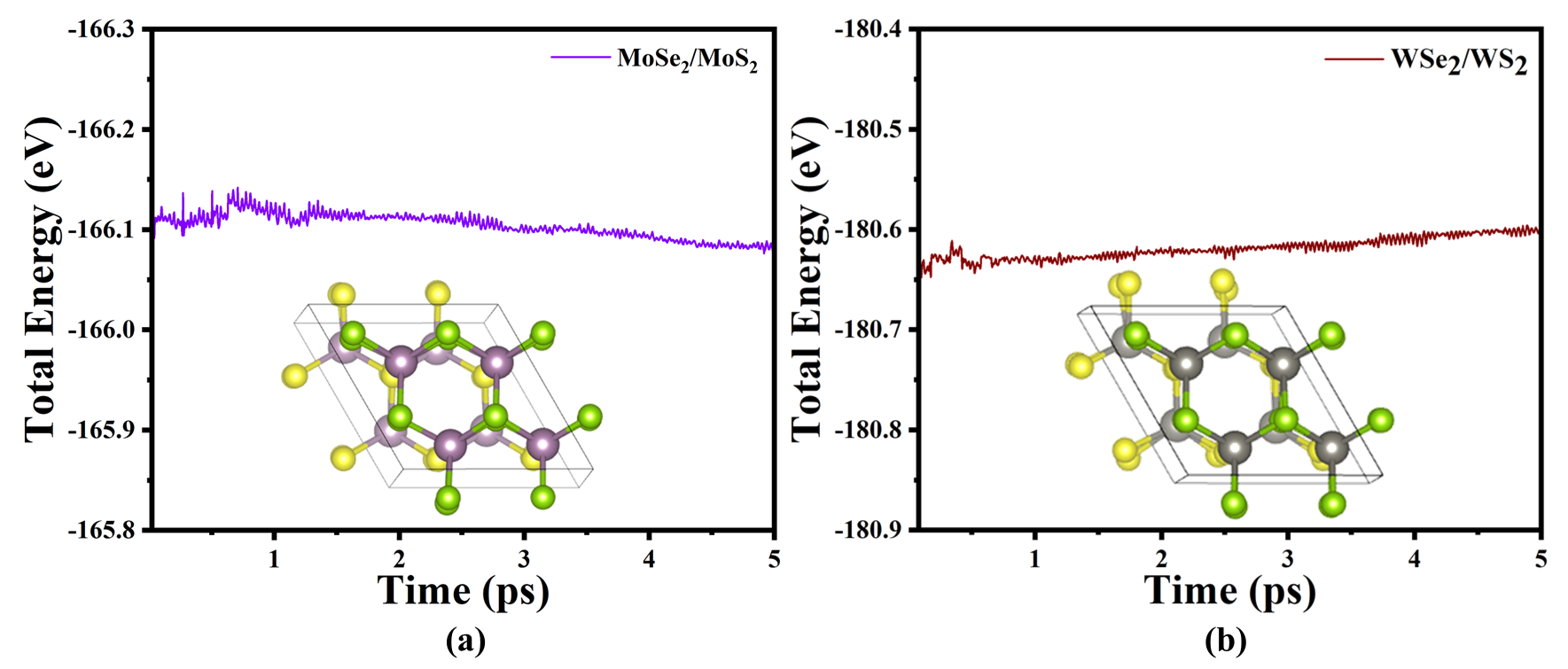}
\par\end{centering}
\caption{\protect\label{fig:md_main}Total energy as a function of time steps
for (a) MoSe\protect\textsubscript{2}/MoS\protect\textsubscript{2}
and (b) WSe\protect\textsubscript{2}/WS\protect\textsubscript{2}
systems obtained from the AIMD simulations at 500K. The resultant
structures are shown in the inset. Here, purple, grey, yellow, and
green spheres represent the Mo, W, S, and Se atoms, respectively.}
\end{figure}

Furthermore, ab-initio molecular dynamics (AIMD) \citep{Kresse1994aimd}
simulations have also been performed to explore the stability of the
considered heterostructures. These simulations were performed on larger
sized 2$\times$2 supercells of the considered heterostructures, as
compared to the DFT calculations that were done on the primitive cells.
The AIMD calculations were carried out in canonical ensemble utilizing
the Nose-Hoover thermostat \citep{nose1984} at 500K for a total duration
of 5000 fs in the steps of 1 fs. The resultant total energy as a function
of time steps is depicted in Fig. \ref{fig:md_main} for the MoSe\textsubscript{2}/MoS\textsubscript{2}
and WSe\textsubscript{2}/WS\textsubscript{2} systems, while for
the rest of the heterostructures, they are shown in Fig. S4 of the
SM. The plotted graphs showed minimal variations
in the total energies, which confirms the stability of the considered
heterostructures at higher temperatures as well. The corresponding
structures obtained during the AIMD simulations are also shown in
the inset. The maximum average changes in the Mo--S, Mo--Se, W--S,
and W--Se bond lengths during the course of the simulations are 0.06 
 {\AA},
0.05 {\AA}, 0.06 {\AA}, and 0.03 {\AA}, respectively, which are quite small. 

\subsubsection{Electronic Properties}

\begin{figure}[H]
\begin{centering}
\includegraphics[scale=0.3]{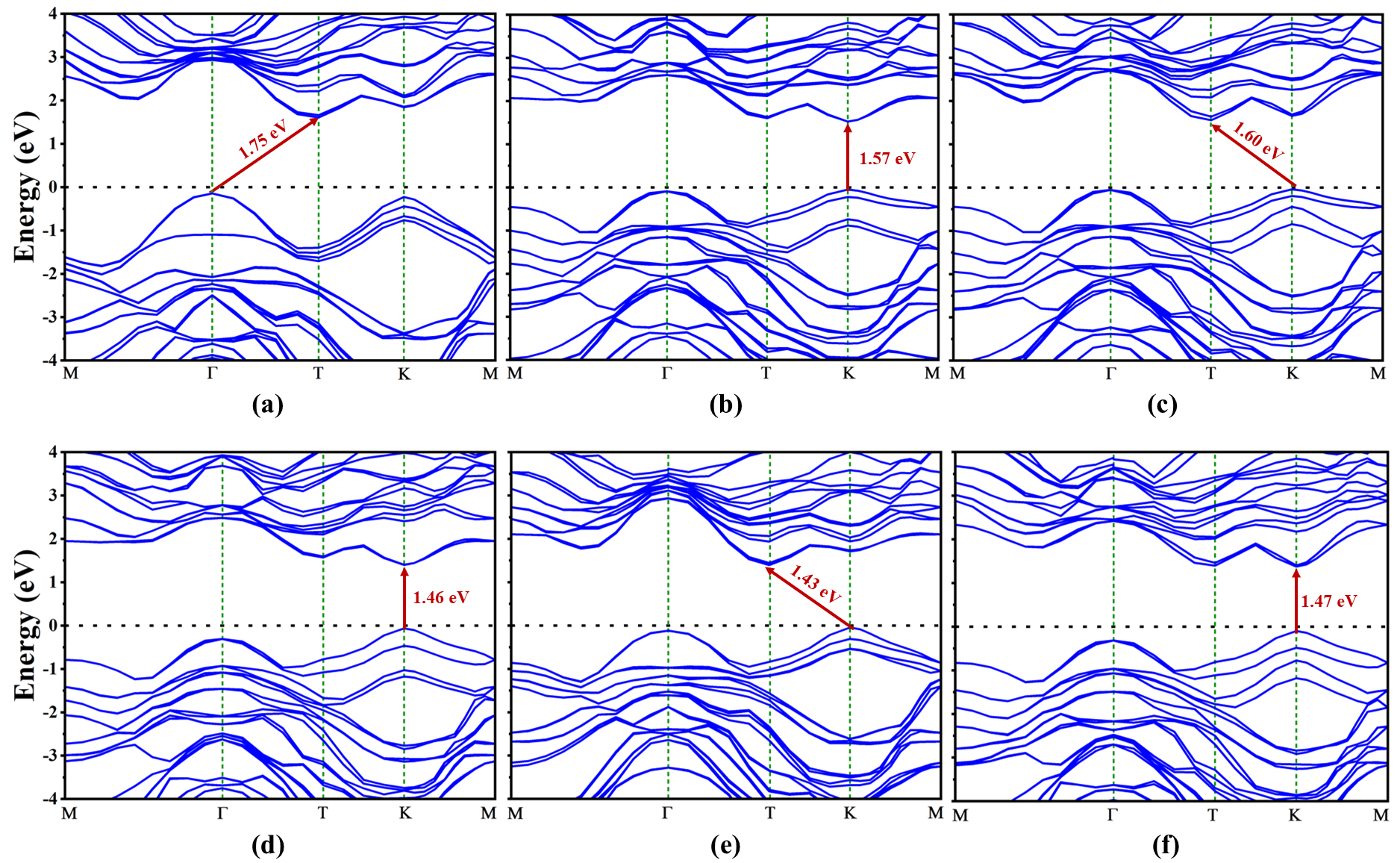}
\par\end{centering}
\caption{\protect\label{fig:bandstruct-HSE06}Calculated band structures of
(a) WS\protect\textsubscript{2}/MoS\protect\textsubscript{2}, (b)
MoSe\protect\textsubscript{2}/MoS\protect\textsubscript{2}, (c)
MoSe\protect\textsubscript{2}/WS\protect\textsubscript{2}, (d) WSe\protect\textsubscript{2}/MoS\protect\textsubscript{2},
(e) WSe\protect\textsubscript{2}/MoSe\protect\textsubscript{2},
and (f) WSe\protect\textsubscript{2}/WS\protect\textsubscript{2}
heterojunctions using the HSE06 functional\textcolor{blue}{.}}
\end{figure}

The electronic band structures of the studied heterostructures were
initially calculated using both the GGA-PBE and HSE06 functionals.
To confirm the spatial separation of charge carriers, we computed
the atomic orbital-projected PBE band structures, shown in Fig. S5
of the SM. These highlight the contributions of
each element's valence orbitals to the energy states. The analysis
reveals that the VBM and CBM belong to different layers: the VBM primarily
originates from the donor layer, while the CBM is composed of the
acceptor layer. This indicates that the photogenerated electrons and
holes are confined to separate layers, in accordance with the type-II
band alignment, confirmed previously for all six heterostructures
at the HSE06 level. For example, in the MoSe\textsubscript{2}/MoS\textsubscript{2}
heterostructure (see Fig. S5(b) of the SM), VBM
is attributed to the MoSe\textsubscript{2} (donor) layer, while contribution
to CBM comes from the MoS\textsubscript{2} (acceptor) layer. The obtained
$E_{g}$ values from the PBE band structures are listed in Table \ref{tab:hetero-electronic-gaps}.
The PBE band structures yield indirect $E_{g}$ of 1.15 eV ($\Gamma\rightarrow$T),
1.09 (K$\rightarrow$T), and 0.95 eV (K$\rightarrow$T), respectively
for the WS\textsubscript{2}/MoS\textsubscript{2} \citep{Liang2014,Amin2015},
MoSe\textsubscript{2}/WS\textsubscript{2} \citep{Amin2016}, and
WSe\textsubscript{2}/MoSe\textsubscript{2} \citep{Hu2016,Amin2015}
heterostructures, consistent with the literature. Some minor differences
in the k-point positions for the CBM are observed, such as in the
WS\textsubscript{2}/MoS\textsubscript{2} heterostructure, where
Amin \textit{et al.} \citep{Amin2015} reported a gap from $\Gamma\rightarrow$K,
while Liang \textit{et al.} \citep{Liang2014} reported VBM at $\Gamma$
and CBM at a k-point in the mid of $\Gamma$ and K, in agreement with
our results. The disagreement with the work of Amin \textit{et al.}
\citep{Amin2015} may arise from our use of non-local vdW corrections,
compared to local corrections used in their study. The direct $E_{g}$
(at K) of 0.93 eV, 0.64 eV, and 0.90 eV for the MoSe\textsubscript{2}/MoS\textsubscript{2},
WSe\textsubscript{2}/MoS\textsubscript{2}, and WSe\textsubscript{2}/WS\textsubscript{2}
heterostructures, respectively, are also in agreement with the previous
studies \citep{komsa2013,Wang2017,Krivosheeva2020,Terrones2013}.
The underestimated PBE band gaps get enhanced using the HSE06 functional
(see $E_{g}$ values presented in Table \ref{tab:hetero-electronic-gaps}).
The HSE06 band structures, presented in Fig.\ref{fig:bandstruct-HSE06}
show that while the conduction band shifts upward, the overall band
dispersion remains unchanged, maintaining the same band gap nature
as in the PBE calculations. For the WS\textsubscript{2}/MoS\textsubscript{2},
MoSe\textsubscript{2}/WS\textsubscript{2}, and WSe\textsubscript{2}/MoSe\textsubscript{2}
heterostructures our HSE06 based $E_{g}$ values of 1.75 eV, 1.60
eV, and 1.43 eV, respectively, are consistent with the corresponding
reported gaps of 1.70 eV \citep{Amin2015}, 1.58 eV \citep{Amin2016},
1.5 eV \citep{Amin2015}. We further observe that, irrespective of
the functional used, the highest occupied valence bands of the heterostructures
are degenerate throughout the k-path M$\rightarrow\Gamma$ and split
midway through $\Gamma\rightarrow$T. When the same calculations are
performed without including the SOC, no such splitting takes place,
therefore, this splitting is clearly due to SOC. This behavior is
seen for all the heterostructures under consideration.

\begin{table}
\caption{\protect\label{tab:hetero-electronic-gaps}Calculated $E_{g}$ values
using the PBE and HSE06 functionals, and $E_{g}^{QP}$ values using
G$_{0}$W$_{0}$ approximation considering 160 unoccupied bands. $E_{g}^{QP}$
values are indirect in nature for all the considered heterostructures
. }

\centering{}%
\begin{tabular}{cccccccc}
\toprule 
\multirow{2}{*}{S.No.} & \multirow{2}{*}{Heterostructures} & \multicolumn{3}{c}{$E_{g}$(eV)} & $E_{g}^{QP}$(eV)  & \multicolumn{2}{c}{$E_{dir}^{QP}$ (eV) }\tabularnewline
\cmidrule{3-8}
 &  & PBE & HSE06 &  & G$_{0}$W$_{0}$ & \ \ \ G$_{0}$W$_{0}$ & \ G$_{0}$W$_{0}$+ SOC\tabularnewline
\midrule
\midrule 
1 & WS\textsubscript{2}/MoS\textsubscript{2} & 1.15 & 1.75 & Indirect & 2.15 & \ \ \ 2.99 & 2.79\tabularnewline
2 & MoSe\textsubscript{2}/MoS\textsubscript{2} & 0.93 & 1.57 & Direct & 1.90 & \ \ \ 2.43 & 2.34\tabularnewline
3 & MoSe\textsubscript{2}/WS\textsubscript{2} & 1.09 & 1.60 & Indirect & 1.94 & \ \ \ 2.60 & 2.44\tabularnewline
4 & WSe\textsubscript{2}/MoS\textsubscript{2} & 0.64 & 1.46 & Direct & 2.08 & \ \ \ 2.34 & 2.14\tabularnewline
5 & WSe\textsubscript{2}/MoSe\textsubscript{2} & 0.95 & 1.43 & Indirect & 2.12 & \ \ \ 2.79 & 2.58\tabularnewline
6 & WSe\textsubscript{2}/WS\textsubscript{2} & 0.90 & 1.47 & Direct & 2.15 & \ \ \ 2.70 & 2.00\tabularnewline
\bottomrule
\end{tabular}
\end{table}

\begin{figure}
\begin{centering}
\includegraphics[scale=0.3]{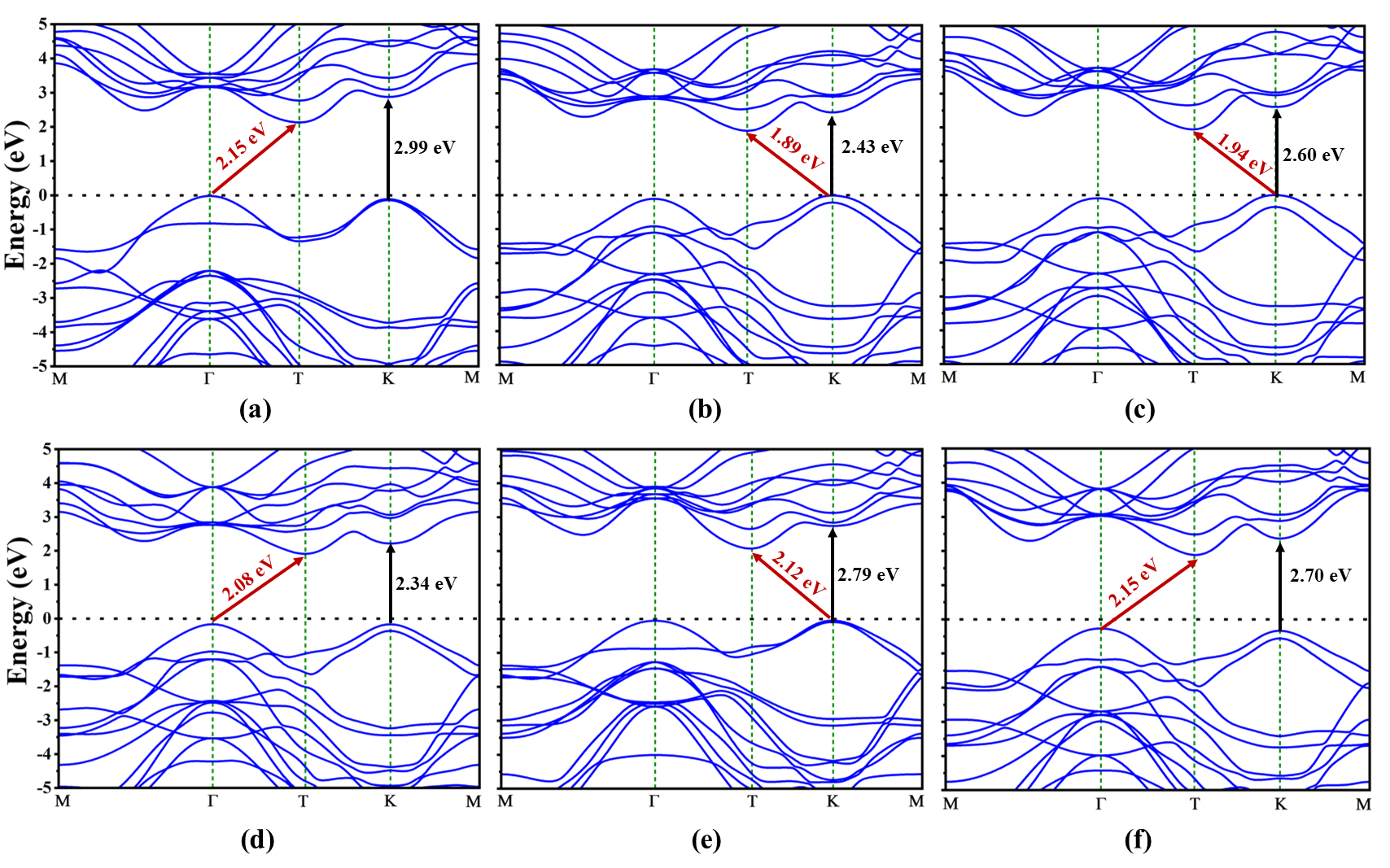}
\par\end{centering}
\caption{\protect\label{fig:bandstrc-GW}Calculated band structures of (a)
WS\protect\textsubscript{2}/MoS\protect\textsubscript{2}, (b) MoSe\protect\textsubscript{2}/MoS\protect\textsubscript{2},
(c) MoSe\protect\textsubscript{2}/WS\protect\textsubscript{2}, (d)
WSe\protect\textsubscript{2}/MoS\protect\textsubscript{2}, (e) WSe\protect\textsubscript{2}/MoSe\protect\textsubscript{2},
and (f) WSe\protect\textsubscript{2}/WS\protect\textsubscript{2}
heterojunctions using G$_{0}$W$_{0}$ approximation.}

\end{figure}

Given the strong correlation effects observed in the QP gaps of the
TMD monolayers, it is likely that these effects also play a significant
role in the corresponding heterostructures. Therefore, we performed
QP band structure calculations using G$_{0}$W$_{0}$ approximation
for all six TMD heterostructures. Initially, 160 unoccupied bands
were considered, and this number was increased to 200 to check the
convergence of the QP band gaps ($E_{g}^{QP}$). The $E_{g}^{QP}$
values using 160 bands are listed in Table \ref{tab:hetero-electronic-gaps},
while those with 200 bands are in Table S3 of the SM.
The results show almost negligible changes in $E_{g}^{QP}$ with 200
bands, so we proceed with 160 unoccupied bands for further study of
G$_{0}$W$_{0}$ band structures for all the six heterostructures.
For the purpose, a procedure identical to the case of monolayers was
adopted and the corresponding Wannierized PBE band structure for WS\textsubscript{2}/MoS\textsubscript{2}
is compared with the original band structure in Fig. S6 of the SM. The Wannier-interpolated G$_{0}$W$_{0}$ band
structures for all the considered heterostructures are depicted in
Fig. \ref{fig:bandstrc-GW}. We found that for MoSe\textsubscript{2}/MoS\textsubscript{2},
WSe\textsubscript{2}/MoS\textsubscript{2}, and WSe\textsubscript{2}/WS\textsubscript{2}
heterostructures which showed direct gaps at the DFT level (PBE and
HSE06), the CBM shifts to a lower energy at the T point in the Brillouin
zone located between $\Gamma$ and K, resulting in indirect gaps.
While the other three heterostructures, namely, WS\textsubscript{2}/MoS\textsubscript{2}
\citep{Liang2014,Amin2015}, MoSe\textsubscript{2}/WS\textsubscript{2}
\citep{Amin2016}, and WSe\textsubscript{2}/MoSe\textsubscript{2}
\citep{Hu2016,Amin2015}, which exhibited indirect band gaps at the
DFT level, retained that nature even after G$_{0}$W$_{0}$ calculations.
In the WSe\textsubscript{2}/MoS\textsubscript{2} heterostructure,
after G$_{0}$W$_{0}$ calculations, the VBM is shifted from K to
$\Gamma$, compared to the HSE06 results. The $E_{g}^{QP}$ values
for all the heterostructures are significantly larger than those obtained
from the DFT calculations, indicating strong electron correlation
effects. For WS\textsubscript{2}/MoS\textsubscript{2} heterostructure,
our calculated $E_{g}^{QP}$ of 2.15 eV is in excellent agreement
with the reported value of 2.10 eV \citep{Amin2015}, along with the
indirect nature of the band gap. Similarly, our indirect $E_{g}^{QP}$
of 2.12 eV for the WSe\textsubscript{2}/MoSe\textsubscript{2} heterostructure
is also consistent with the literature \citep{Torun2018}. In the
G$_{0}$W$_{0}$ bands of all heterostructures, a direct band gap
of higher energy is present at the K high symmtery point, as is clear
from Fig. \ref{fig:bandstrc-GW}, and the corresponding $E_{dir}^{QP}$
values are also given in Table \ref{tab:hetero-electronic-gaps}.
These direct gaps are crucial for studying the optical response of
the heterostructures discussed later in detail in section \ref{subsec:Optical-Properties}.
The SOC effects are included in $E_{dir}^{QP}$ (see Table \ref{tab:hetero-electronic-gaps}),
similar to the case of monolayers, by applying a rigid shift to the
VBM and CBM, based on the shifts obtained at the GGA-PBE level.

Heterojunction formation leads to a reduction in both the DFT and
G$_{0}$W$_{0}$ band gaps compared to the isolated TMD monolayers.
For example, the $E_{g}$ ($E_{g}^{QP}$) for isolated WS\textsubscript{2}
and WSe\textsubscript{2} monolayers are 1.96 eV (2.76 eV) and 1.65
eV (2.17 eV), respectively. After combining these layers into the
WSe\textsubscript{2}/WS\textsubscript{2} heterostructure, the $E_{g}$
($E_{g}^{QP}$) decrease to 1.47 eV (2.15 eV). This reduction in band
gaps due to heterojunction formation is advantageous for covering
a broader solar spectrum. 

\begin{figure}[H]
\begin{centering}
\includegraphics[scale=0.5]{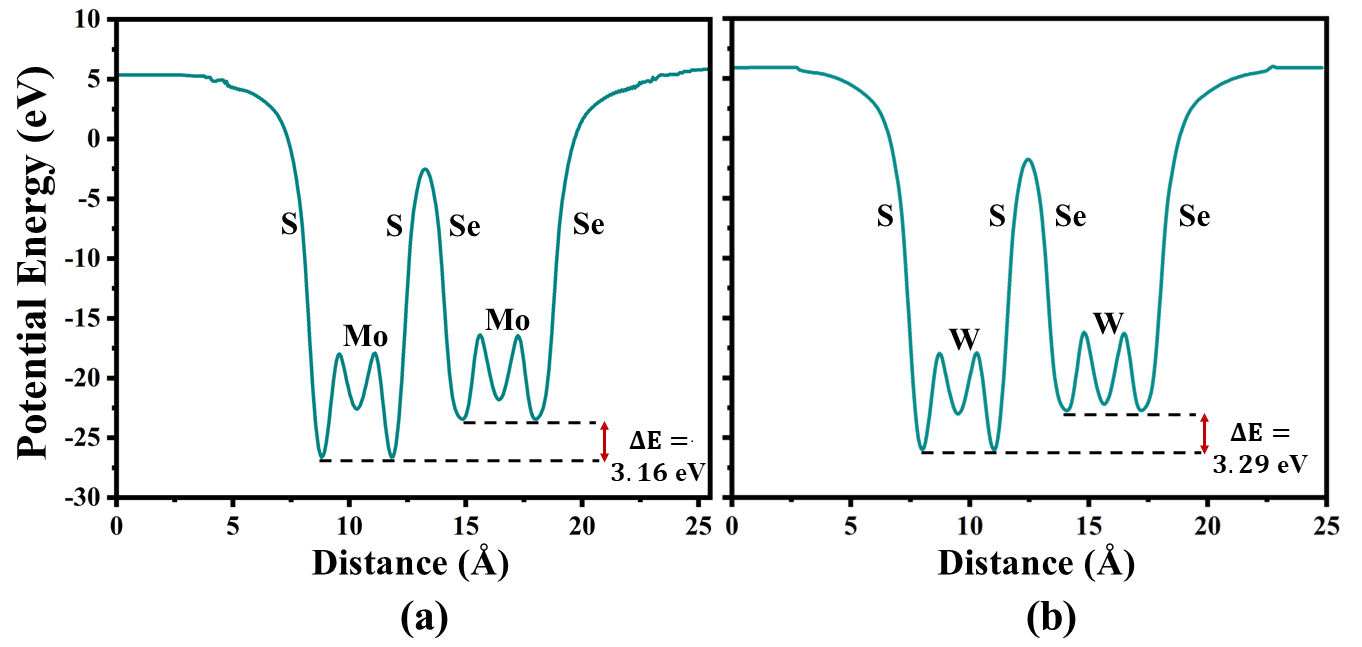}
\par\end{centering}
\caption{\protect\label{fig:potential-energy}Electrostatic potential energy
as a function of distance along z-direction for (a) MoSe\protect\textsubscript{2}/MoS\protect\textsubscript{2}
and (b) WSe\protect\textsubscript{2}/WS\protect\textsubscript{2}
heterojunctions. $\Delta E$ denotes the difference in average electrostatic
potential energy across the layers.}
\end{figure}

The built-in electric field/voltage in a solar cell is an important
quantity determining its operation by playing a key role in the separation
and movement of charge carriers. Due to the horizontal mirror asymmetry
of the considered heterostructures, they will have in-built electric
fields in the $z$-direction. The electrostatic potential energy as
a function of distance along the $z$-direction is presented in Fig.\ref{fig:potential-energy}
for the MoSe\textsubscript{2}/MoS\textsubscript{2} and WSe\textsubscript{2}/WS\textsubscript{2}
heterostructures, while for the rest of the four heterostructures,
it is shown in Fig. S7 of the SM. The asymmetric
potential energy across the two layers gives rise to an in-built electric
field ($\boldsymbol{E_{in}}$) across each of the considered heterostructures.
The difference in average potential energy across the two layers,
denoted by $\Delta E$ (see Table \ref{tab:effective=000020mass}),
is 3.16 eV and 3.29 eV in the case of MoSe\textsubscript{2}/MoS\textsubscript{2}
and WSe\textsubscript{2}/WS\textsubscript{2} heterostructures, respectively
(see Fig.\ref{fig:potential-energy}). For the other four heterostructures,
WS\textsubscript{2}/MoS\textsubscript{2}, MoSe\textsubscript{2}/WS\textsubscript{2},
WSe\textsubscript{2}/MoS\textsubscript{2}, and WSe\textsubscript{2}/MoSe\textsubscript{2},
$\Delta E$ of 0.08 eV, 3.31 eV, 3.17 eV, and 0.06 eV are obtained,
respectively. The quite small $\Delta E$ values in the case of WS\textsubscript{2}/MoS\textsubscript{2}
and WSe\textsubscript{2}/MoSe\textsubscript{2} heterostructures
are due to the presence of similar chalcogen atoms in both the constituent
monolayers, whereas, comparatively large $\Delta E$ values in case
of other heterostructures arises due to the electronegativity differences
(S > Se). Using the $\Delta E$ values, $\boldsymbol{E_{in}}$ is
calculated, and their values (see Table \ref{tab:effective=000020mass}),
as expected, are significant only for those four heterostructures
(MoSe\textsubscript{2}/MoS\textsubscript{2}, MoSe\textsubscript{2}/WS\textsubscript{2},
WSe\textsubscript{2}/MoS\textsubscript{2}, and WSe\textsubscript{2}/WS\textsubscript{2,})
for which $\Delta E$ is large. This intrinsic electric field helps
to spatially separate the photogenerated charge carriers because,
under its influence, the electrons and holes move in opposite directions.
This separation of charges is crucial for the generation of an electric
current when an external circuit is connected to the solar cell. In
a sense, the built-in electric field provides the initial push for
the flow of electrons, thus enabling the conversion of light energy
into electrical energy. Clearly, higher built-in electric fields will
lead to highly efficient solar cell devices. 

\subsubsection{Effective masses of charge carriers}

The effective masses of the charge carriers of a semiconductor determine
their mobility, and, thus the performance of a solar cell composed
of it. In this work, we have computed the effective masses of both
the electrons ($m_{e}^{*}$) and holes ($m_{h}^{*}$) by employing
the parabolic approximation around the CBM and VBM, respectively.
The following equation is used for calculating the elements of the
effective mass tensor: 
\begin{equation}
m_{ij}^{*}({\bf k})=\hbar^{2}\left(\frac{\partial^{2}(E({\bf k}))}{\partial k_{i}\partial k_{j}}\right)_{{\bf k}={\bf K}_{0}}^{-1}m_{0}
\end{equation}

where $\hbar$ is the reduced Planck's constant, ${\bf k}$ is a vector
in the first Brillouin zone, and $E({\bf k})$ is the corresponding
valence/conduction band Kohn-Sham eigenvalue. ${\bf K}_{0}$ represents
the band extremum location in the band structure of all the considered
heterostructures, $i$, $j$ are Cartesian directions, and $m_{0}$
denotes the free electron mass. Because of the isotropic nature of
these materials, there is only one unique nonzero ``in plane'' element
of the effective-mass tensor of the holes and electrons of each heterostructure,
which we report in Table \ref{tab:effective=000020mass}. The effective
masses are calculated at the CBM and VBM along specific $k$ directions
specified in the parenthesis next to them. For instance, in the WSe\textsubscript{2}/MoSe\textsubscript{2}
heterostructure, $m_{e}^{*}$ ( $m_{h}^{*}$) is calculated along
the high symmetry directions T$\rightarrow\Gamma$ (K$\rightarrow$T)
and T$\rightarrow$K (K$\rightarrow$M). Furthermore, we have compared
these effective masses with those of involved monolayers, provided
in Table S2 of the SM \citep{supporting}. We note that the calculated
values of the effective masses in most of the considered heterostructures
are close to their values in the constituent monolayers, except for
WS\textsubscript{2}/MoS\textsubscript{2}, for which $m_{h}^{*}$
for the heterostructure is comparatively larger. Smaller $m_{e}^{*}$
and $m_{h}^{*}$ values indicate higher carrier mobilities, which
improve both carrier current and photogenerated charge separation.
Ideally, $m_{e}^{*}$ and $m_{h}^{*}$ should be of similar magnitudes
to ensure balanced carrier mobilities in opposite directions, reducing
electron-hole recombination rate. We found that in WSe\textsubscript{2}/MoS\textsubscript{2},
WSe\textsubscript{2}/MoSe\textsubscript{2}, and WSe\textsubscript{2}/WS\textsubscript{2}
heterostructures, $m_{e}^{*}$ and $m_{h}^{*}$ are smaller compared
to those for the other heterostructures and nearly equal, suggesting
better solar cell performance.

\begin{table}[H]
\caption{\protect\label{tab:effective=000020mass}The difference in average
electrostatic potential energy across the layers ($\Delta E$), in-built
electric field ($\boldsymbol{E_{in}}$), and the computed effective
masses of electrons ($m_{e}^{*}$) and holes ($m_{h}^{*}$) in the
units of the free-electron mass, $m_{0}$ for the heterojunctions
under consideration.}

\begin{centering}
\begin{tabular}{cccccccc}
\toprule 
\multirow{1}{*}{S.No.} & \multirow{1}{*}{Heterojunction} & $\Delta E$(eV) & $\boldsymbol{E_{in}}$(V/\AA) & \multicolumn{2}{c}{$m_{e}^{*}$($m_{0}$)} & \multicolumn{2}{c}{$m_{h}^{*}(m_{0})$}\tabularnewline
\midrule
\midrule 
1 & WS\textsubscript{2}/MoS\textsubscript{2} & 0.08 & 0.01 & 0.53(T$\rightarrow\Gamma$) & 0.60(T$\rightarrow$K) & 0.86($\Gamma\rightarrow$M) & 0.86($\Gamma\rightarrow$T)\tabularnewline
2 & MoSe\textsubscript{2}/MoS\textsubscript{2} & 3.16 & 0.34 & 0.45(K$\rightarrow$T) & 0.49(K$\rightarrow$M) & 0.70(K$\rightarrow$T) & 0.89(K$\rightarrow$M)\tabularnewline
3 & MoSe\textsubscript{2}/WS\textsubscript{2} & 3.31 & 0.36 & 0.48(T$\rightarrow\Gamma$) & 0.53(T$\rightarrow$K) & 0.69(K$\rightarrow$T) & 0.78(K$\rightarrow$M)\tabularnewline
4 & WSe\textsubscript{2}/MoS\textsubscript{2} & 3.17 & 0.34 & 0.43(K$\rightarrow$T) & 0.46(K$\rightarrow$M) & 0.43(K$\rightarrow$T) & 0.48(K$\rightarrow$M)\tabularnewline
5 & WSe\textsubscript{2}/MoSe\textsubscript{2} & 0.06 & 0.01 & 0.47(T$\rightarrow\Gamma$) & 0.49(T$\rightarrow$K) & 0.39(K$\rightarrow$T) & 0.44(K$\rightarrow$M)\tabularnewline
6 & WSe\textsubscript{2}/WS\textsubscript{2} & 3.29 & 0.36 & 0.36(K$\rightarrow$T) & 0.39(K$\rightarrow$M) & 0.43(K$\rightarrow$T) & 0.48(K$\rightarrow$M)\tabularnewline
\bottomrule
\end{tabular}
\par\end{centering}
\end{table}

\subsubsection{\protect\label{subsec:Optical-Properties}Optical Properties}

\begin{figure}[H]
\begin{centering}
\includegraphics[scale=0.3]{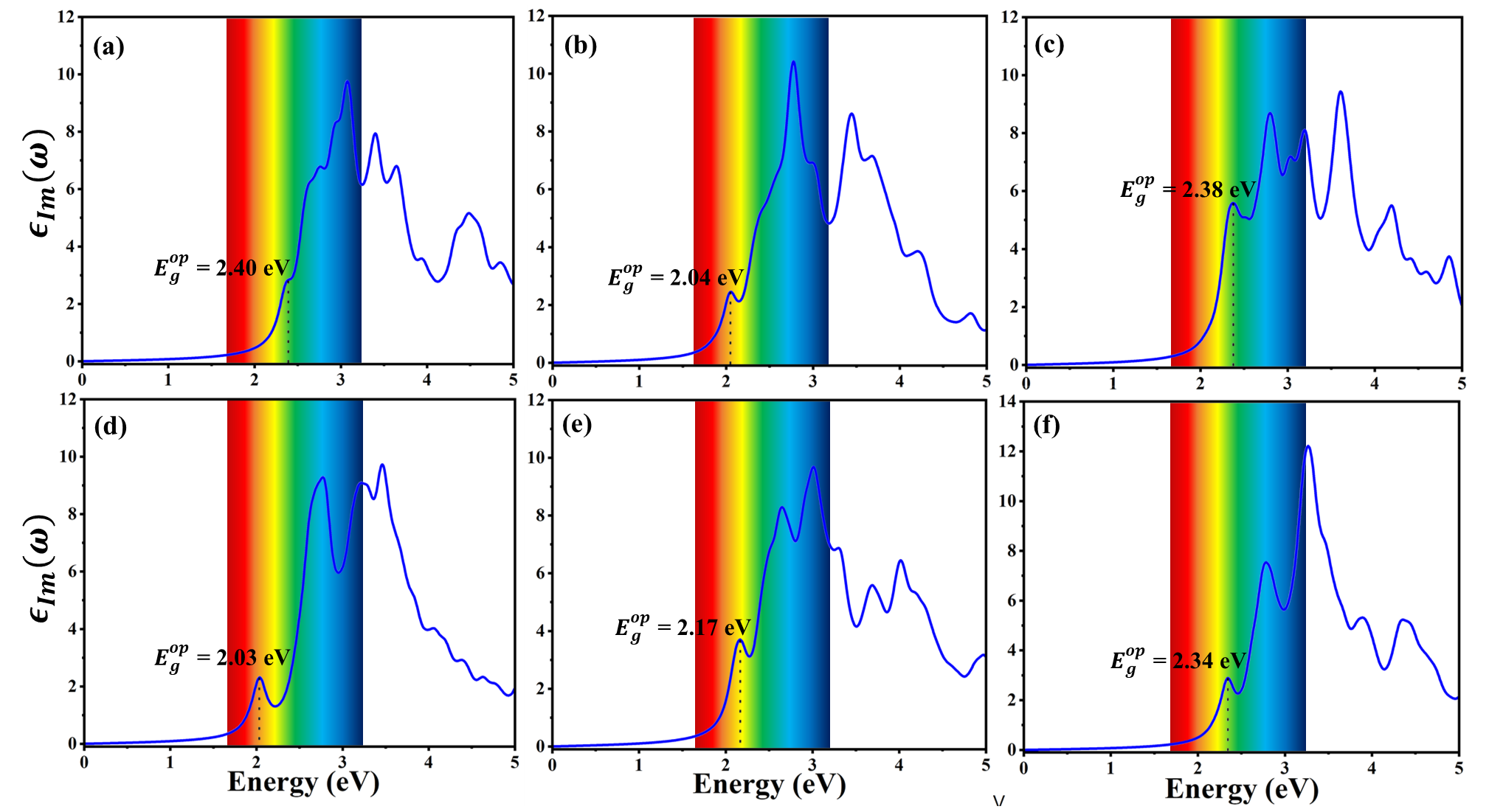}
\par\end{centering}
\caption{\protect\label{fig:BSE-optical-hetero}Imaginary part of the dielectric
function, $\epsilon_{Im}(\omega)$ as a function of the energy of
the incident photon for (a) WS\protect\textsubscript{2}/MoS\protect\textsubscript{2},
(b) MoSe\protect\textsubscript{2}/MoS\protect\textsubscript{2},
(c) MoSe\protect\textsubscript{2}/WS\protect\textsubscript{2}, (d)
WSe\protect\textsubscript{2}/MoS\protect\textsubscript{2}, (e) WSe\protect\textsubscript{2}/MoSe\protect\textsubscript{2},
and (f) WSe\protect\textsubscript{2}/WS\protect\textsubscript{2 }heterostructures
calculated using BSE. }
\end{figure}

After exploring the electronic properties, our study focuses on the
optical response of the heterostructures under consideration. For
the study of optical properties, we have employed BSE as given by
Eq.\ref{eq:BSE-eq}. Similar to monolayers, $\epsilon_{Im}(\omega)$
is plotted against the incident photon energy for each heterostructure,
as depicted in Fig. \ref{fig:BSE-optical-hetero}. It is to be noted
that the nature of $E_{g}^{QP}$ computed using the G$_{0}$W$_{0}$
approximation (on the top of which the BSE calculations are performed)
is indirect for all the six heterostructures. As a result, the optical
transitions corresponding to these gaps are not reflected in the optical
absorption spectra, and the transitions corresponding to $E_{dir}^{QP}$
(at K, shown by an arrow in Fig. \ref{fig:bandstrc-GW}) dominate
the computed BSE spectra presented in Fig. \ref{fig:BSE-optical-hetero}.
The first optically active peak arises by excitonic transitions, and
corresponds to the optical gaps ($E_{g}^{op}$) listed in Table \ref{tab:Hetero-optical-gaps,}.
These $E_{g}^{op}$ values lie in the visible spectrum, with the coverage
of about half of the visible range (highlighted in the figure), which
is the focus of this work because of its possible solar-cell applications.
For the MoSe\textsubscript{2}/MoS\textsubscript{2} heterostructure,
$E_{g}^{op}$ of 2.04 eV is close to the reported photoluminescence
(PL) peak of 1.89 eV \citep{Ceballos2014}, validating our calculations.
Further, we added the effect of SOC in all the $E_{g}^{op}$ values
as done in the $E_{g}^{QP}$ values of the studied heterostructures.
The resultant $E_{g}^{op}$ values at the BSE+SOC level are also listed
in the same table. Remarkably, the SOC-corrected BSE gives $E_{g}^{op}$
of 1.95 eV for the MoSe\textsubscript{2}/MoS\textsubscript{2} heterostructure
aligns even better with the reported PL peak \citep{Ceballos2014}.
The $E_{ex}$ is also calculated for each of the heterostructure,
and the obtained values are in the range of 0.22 eV -- 0.62 eV. We
observe that in most of the cases, $E_{ex}$ is smaller in heterostructures
as compared to the constituent monolayers, such as 0.39 eV for the
MoSe\textsubscript{2}/MoS\textsubscript{2} heterostructure versus
0.46 eV for the MoSe\textsubscript{2} and MoS\textsubscript{2} monolayers.
The lower $E_{ex}$ results from increased screening due to heterostructure
formation, leading to weaker exciton binding, and thus easier separation
of the photogenerated charge carriers.

\begin{table}[H]

\caption{\protect\label{tab:Hetero-optical-gaps,}Calculated optical gaps,
$E_{g}^{op}$ from the BSE calculations and the corresponding exciton
binding energies, $E_{ex}$ for the TMD heterostructures under consideration.}

\centering{}%
\begin{tabular}{ccccc}
\toprule 
\multirow{2}{*}{S.No.} & \multirow{2}{*}{Heterostructures} & \multicolumn{2}{c}{$E_{g}^{op}$ (eV)} & $E_{ex}$ (eV)\tabularnewline
\cmidrule{3-5}
 &  & BSE & BSE+SOC & \tabularnewline
\midrule
\midrule 
1 & WS\textsubscript{2}/MoS\textsubscript{2} & 2.40 & 2.20 & 0.59\tabularnewline
2 & MoSe\textsubscript{2}/MoS\textsubscript{2} & 2.04 & 1.95 & 0.39\tabularnewline
3 & MoSe\textsubscript{2}/WS\textsubscript{2} & 2.38 & 2.22 & 0.22\tabularnewline
4 & WSe\textsubscript{2}/MoS\textsubscript{2} & 2.03 & 1.83 & 0.31\tabularnewline
5 & WSe\textsubscript{2}/MoSe\textsubscript{2} & 2.17 & 1.96 & 0.62\tabularnewline
6 & WSe\textsubscript{2}/WS\textsubscript{2} & 2.34 & 1.64 & 0.36\tabularnewline
\bottomrule
\end{tabular}
\end{table}

\subsubsection{Power Conversion Efficiencies (PCE)}

The power-conversion efficiency (PCE) of a solar device, as its name
suggests, is a measure of how efficiently it converts the solar energy
into the electrical energy. To calculate PCE, a method proposed by
Scharber \emph{et al.} \citep{scharber2006} has been widely used,
according to which the maximum PCE under the 100\% external quantum
efficiency (EQE) can be determined as

\begin{equation}
PCE=\frac{V_{oc}J_{sc}\beta_{FF}}{P_{solar}}=\frac{0.65(E_{g}^{d}-\Delta E_{c}-0.3)\stackrel[E_{g}^{d}]{\infty}{\int}\frac{P(\hbar\omega)}{\hbar\omega}d\omega}{\stackrel[0]{\infty}{\int}P(\hbar\omega)d\omega},\label{eq:PCE}
\end{equation}

where 0.65 is the band-fill factor denoted by $\beta_{FF}$, $V_{oc}=(E_{g}^{d}-\Delta E_{c}-0.3)$
is an estimate of the open circuit voltage, with $E_{g}^{d}$ and
$\Delta E_{c}$ representing the band gap of donor and the conduction
band offset between the donor and acceptor semiconductors, respectively,
$P(\hbar\omega)$, expressed in the units Wm$^{-2}$eV$^{-1}$, is
the AM1.5 solar energy flux for energy $\hbar\omega$ of the photon.
Further, $J_{sc}$ is the short circuit current density evaluated
as the integral $J_{sc}=\stackrel[E_{g}^{d}]{\infty}{\int}\frac{P(\hbar\omega)}{\hbar\omega}d\omega$
, under the assumption of 100\% EQE. The integral in the denominator
$\stackrel[0]{\infty}{\int}P(\hbar\omega)d\omega$ is the total power
of the incident solar radiation. To date, reported PCEs for various
materials in numerous studies were calculated at the DFT level only,
using the hybrid functionals \citep{Guo2014,Liang2018,WuKai2019}.
Following this, we calculated PCEs using the HSE06 functional and
compared our results with those in the literature for different heterojunction
solar cells. Additionally, we also computed PCE for the TMD heterostructures
at the G$_{0}$W$_{0}$ level for completeness. To summarize, we compute
PCE in the following three ways: (a) at the DFT level using the HSE06
functional, (b) at the G$_{0}$W$_{0}$ level without SOC, and (c)
at the G$_{0}$W$_{0}$ level with SOC. For the latter two methods,
we used $E_{g}^{op}$ from BSE as $E_{g}^{d}$, instead of $E_{g}^{QP}$,
because $E_{g}^{op}$ calculated from BSE more accurately reflects
experimental values due to its inclusion of electron-hole attraction
effects. Further, the BSE optical gap is independent of the vacuum
used in calculations, unlike the GW QP gap \citep{komsa2013}. The
resulting PCEs calculated using Eq. \ref{eq:PCE} are summarized in
Table \ref{tab:PCE}, with the required intermediate parameters ($E_{g}^{d}$,
$\Delta E_{c}$, $V_{oc}$, and $J_{sc}$) provided in Tables S4,
S5, and S6 of the SM. PCE is not calculated at
the G$_{0}$W$_{0}$ level for MoSe\textsubscript{2}/WS\textsubscript{2}
because of its failure to show type II band alignment, as discussed
earlier. We obtained PCEs in the ranges of 12.93\% -- 19.25\%, 6.54\%
-- 14.79\% , and 7.48\% -- 15.25\% for the considered heterostructures
using HSE06, G$_{0}$W$_{0}$, and G$_{0}$W$_{0}$+ SOC, respectively.
For each heterostructure, the PCE order is HSE06 > G$_{0}$W$_{0}$+
SOC > G$_{0}$W$_{0}$, which is reverse of the band gap order. This
is because increasing the donor band gap narrows the range of the
solar spectrum coverage during absorption, because in that case only
photons of energies higher than the gap will be useful. From Eq. \ref{eq:PCE},
it is obvious that in order to have high values of PCE, both $\Delta E_{c}$
and $E_{g}^{d}$ should be small, because lower $\Delta E_{c}$ leads
to higher $V_{oc}$, while a lower $E_{g}^{d}$ gives rise to a higher
$J_{sc}$. It is to be noted that at the HSE06 level, a minimum value
of both $\Delta E_{c}$(0.11 eV) and $E_{g}^{d}$(1.65 eV) for the
WSe\textsubscript{2}/WS\textsubscript{2} heterostructure, resulted
in maximum PCE of 19.25\%. At the G$_{0}$W$_{0}$ and G$_{0}$W$_{0}$+
SOC levels, both $\Delta E_{c}$ and $E_{g}^{d}$ are minimum for
the MoSe\textsubscript{2}/MoS\textsubscript{2} heterostructure,
giving maximum PCEs of 14.79\%, and 15.25\%, respectively. If we compare
the cases of WSe\textsubscript{2}/MoS\textsubscript{2}, WSe\textsubscript{2}/MoSe\textsubscript{2},
and WSe\textsubscript{2}/WS\textsubscript{2} structures, where the
donor semiconductor is the same (WSe\textsubscript{2}), the highest
PCE from all the three methods corresponds to the lowest $\Delta E_{c}$
and vice-versa. It is worth mentioning that our calculated PCEs are
significantly higher than those reported in experimental studies \citep{Ming-Yang2016}.
For instance, experimental PCEs for the WSe\textsubscript{2}/MoS\textsubscript{2}
(0.2\%) and WSe\textsubscript{2}/MoSe\textsubscript{2} (0.12\%)
heterostructures are much lower because they were measured under EQE
of 1.5\% and 1.2\%, respectively \citep{Furchi2014,Gong2015}. However,
it is not possible to calculate EQE theoretically, therefore the PCEs
are computed using Eq. \ref{eq:PCE} which assumes 100\% EQE. These
theoretical PCEs computed assuming ideal conditions provide insights
into the comparative performance of materials for solar energy harvesting,
guiding experimentalists in selecting materials for solar cell design.
Among the studied heterostructures, MoSe\textsubscript{2}/MoS\textsubscript{2}
and WSe\textsubscript{2}/WS\textsubscript{2} are most preferable
candidates for heterojunction solar cells due to their maximum PCE
of 18.64\% and 19.25\% (at the HSE06 level), respectively. In table
\ref{tab:PCE_data} we list the PCEs of some recently proposed heterojunction
solar cell devices, calculated at the DFT level using hybrid functional,
and find that the highest PCE (19.25\%) among our heterostructures
is: (a) larger than most of the values reported in the table, and
(b) very close to the highest reported values.

\begin{table}[H]
\centering{}\caption{\protect\label{tab:PCE}Calculated power conversion efficiencies (PCE)
of the heterostructures under consideration.}
\begin{tabular}{ccccc}
\toprule 
\multirow{2}{*}{S.No.} & \multirow{2}{*}{Heterostructures} & \multicolumn{3}{c}{PCE (\%)}\tabularnewline
\cmidrule{3-5}
 &  & HSE06 & G$_{0}$W$_{0}$ & G$_{0}$W$_{0}$+ SOC\tabularnewline
\midrule
\midrule 
1 & WS\textsubscript{2}/MoS\textsubscript{2} & 12.93 & 9.40 & 12.27\tabularnewline
2 & MoSe\textsubscript{2}/MoS\textsubscript{2} & 18.64 & 14.79 & 15.25\tabularnewline
3 & MoSe\textsubscript{2}/WS\textsubscript{2} & 17.79 &  & \tabularnewline
4 & WSe\textsubscript{2}/MoS\textsubscript{2} & 14.35 & 6.54 & 7.48\tabularnewline
5 & WSe\textsubscript{2}/MoSe\textsubscript{2} & 16.15 & 9.30 & 11.82\tabularnewline
6 & WSe\textsubscript{2}/WS\textsubscript{2} & 19.25 & 9.65 & 12.00\tabularnewline
\bottomrule
\end{tabular}
\end{table}

\begin{table}[H]
\centering{}\caption{\protect\label{tab:PCE_data} Power conversion efficiencies (PCE)
of some heterojunction solar cells proposed in the literature.}
\begin{tabular}{ccc}
\toprule 
Heterojuction solar cells & PCE (\%) & References\tabularnewline
\midrule 
Phosphorene/MoS\textsubscript{2} heterobilayer & 11.5-17.5 & \citep{Guo2014}\tabularnewline
MoS\textsubscript{2}/bilayer phosphorene & 16-18 & \citep{Dai2014}\tabularnewline
CBN/PCBM heterojunctions & 10-20 & \citep{Bernardi2012}\tabularnewline
$\alpha$-As/$\alpha$-AsP bilayer & 21.3 & \citep{Maoyuliang2022}\tabularnewline
Graphene/GaAs & 18.5 & \citep{LI2015}\tabularnewline
MoS\textsubscript{2}/$\psi$-phosphorene & 20.26 & \citep{Wangpsi2017}\tabularnewline
TiNF/TiNBr & 18 & \citep{Liang2018}\tabularnewline
TiNCl/TiNBr & 19 & \citep{Liang2018}\tabularnewline
TiNF/TiNCl & 22 & \citep{Liang2018}\tabularnewline
Te/MoS\textsubscript{2} & 10.6 & \citep{WuKai2019}\tabularnewline
Te/MoSe\textsubscript{2} & 17.5 & \citep{WuKai2019}\tabularnewline
Te/WS\textsubscript{2} & 18.8 & \citep{WuKai2019}\tabularnewline
Te/WSe\textsubscript{2} & 18.7 & \citep{WuKai2019}\tabularnewline
Te/MoTe\textsubscript{2} & 20.1 & \citep{WuKai2019}\tabularnewline
Te/WTe\textsubscript{2} & 22.5 & \citep{WuKai2019}\tabularnewline
WS\textsubscript{2}/MoS\textsubscript{2} & 12.93 & This study\tabularnewline
MoSe\textsubscript{2}/MoS\textsubscript{2} & 18.64 & This study\tabularnewline
MoSe\textsubscript{2}/WS\textsubscript{2} & 17.79 & This study\tabularnewline
WSe\textsubscript{2}/MoS\textsubscript{2} & 14.35 & This study\tabularnewline
WSe\textsubscript{2}/MoSe\textsubscript{2} & 16.15 & This study\tabularnewline
WSe\textsubscript{2}/WS2 & 19.25 & This study\tabularnewline
\bottomrule
\end{tabular}
\end{table}

\subsubsection{Prediction of photocatalytic activity}

After thoroughly investigating the photovoltaics related properties
of the considered TMD heterostructures, here we briefly discuss their
photocatalytic properties for hydrogen generation. Photocatalysis
which relies on the photogenerated electrons and holes, is an efficient
approach for hydrogen evolution through water splitting. This process
involves oxidation and reduction half-cell reactions, with the overall
redox reaction represented as:
\begin{equation}
2H_{2}O\ \ \overset{h\nu>1.23eV}{\rightleftharpoons}\ O_{2}(g)+H_{2}(g)\label{eq:redox-reaction}
\end{equation}

\begin{figure}[H]
\begin{centering}
\includegraphics[scale=0.5]{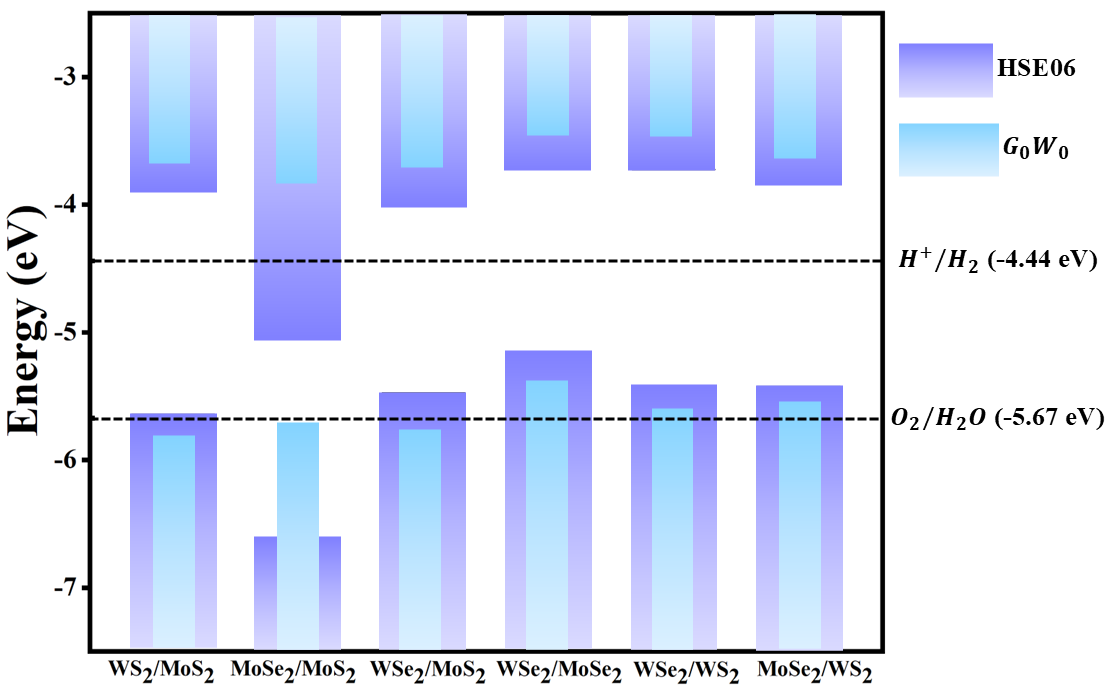}
\par\end{centering}
\caption{\protect\label{alignment-redox}Band edge positions of the considered
TMD heterostructures with respect to vacuum at the HSE06 and G$_{0}$W$_{0}$
level. The redox potential levels are also shown by dashed lines.}
\end{figure}

Experimental data have shown that the equilibrium potentials for the
reduction (H$^{+}/$H$_{2}$) and oxidation (O$_{2}$/H$_{2}$O) reactions
under standard conditions are -4.44 V and -5.67 V, respectively \citep{Gratzel2001}.
Consequently, the potential difference for the corresponding redox
reaction, i.e., $V_{H^{+}/H_{2}}-V_{O_{2}/H_{2}O}$ equals 1.23 V.
This indicates that for a material to serve as an effective photocatalyst,
it must possess a minimum band gap of 1.23 eV. Notably, all the six
TMD heterostructures investigated, meet this band gap criterion according
to both the HSE06 and G$_{0}$W$_{0}$ results. Therefore, all six
TMD heterostructures, qualify as photocatalysts for water-splitting
according to the band gap criteria. Additionally, according to the
Goldilocks principle, an efficient photocatalyst should possess low
recombination rate for which type II band alignment is preferable
\citep{Li2013}. Therefore, MoSe\textsubscript{2}/WS\textsubscript{2}
will not be efficient as a photocatalyst because of its predicted
type I band alignment obtained from G$_{0}$W$_{0}$ approximation.
Furthermore, it is essential to note that while the minimum band gap
criterion is necessary, it is not a sufficient condition for a material
to be an efficient photocatalyst. Another crucial factor is the band
alignment, referring to the alignment of CBM and VBM edges with the
reduction and oxidation potentials of the overall redox reaction (Eq.
\ref{eq:redox-reaction}). According to the band alignment condition,
CBM (VBM) energy level should be less (more) negative than the reduction
(oxidation) energy level. This necessitates that both the potential
levels of the redox reaction should lie within the CBM and VBM levels,
in addition to having a gap larger than 1.23 eV. Therefore, to check
the band alignment for photocatalysis, we have calculated band edges,
$E_{CBM}$ and $E_{VBM}$, at the HSE06 and G$_{0}$W$_{0}$ level,
provided in Table \ref{tab:hetero-bandedges} and also depicted in
Fig. \ref{alignment-redox}. First discussing the HSE06 results, we
observe that none of the considered heterostructures, satisfy the
band alignment criteria. Except MoSe\textsubscript{2}/MoS\textsubscript{2},
$E_{CBM}$ in other five heterostructures meets the band gap criterion
as it is less negative than the reduction potential and thus possesses
strong reducing ability, however, $E_{VBM}$ fails to meet the condition.
In the MoSe\textsubscript{2}/MoS\textsubscript{2}
heterostructure, $E_{VBM}$ aligns with the oxidation level according
to the given criterion and possesses good oxidizing ability, but,
$E_{CBM}$ is more negative than the reduction potential. Next, considering
the band edge positions obtained at the G$_{0}$W$_{0}$ level, we
find that three heterostructures, namely, WS\textsubscript{2}/MoS\textsubscript{2},
MoSe\textsubscript{2}/MoS\textsubscript{2},
and WSe\textsubscript{2}/MoS\textsubscript{2},
meet the band alignment criteria and possess both the reducing and
oxidizing capabilities. The rest three heterostructures, partially
satisfy the criteria, i.e., from the CBM side only. Therefore, from
our G$_{0}$W$_{0}$ calculations, we find that the WS\textsubscript{2}/MoS\textsubscript{2},
MoSe\textsubscript{2}/MoS\textsubscript{2},
and WSe\textsubscript{2}/MoS\textsubscript{2}
heterostructures possess the following three properties: (a) band
gap > 1.23 eV, (b) type II band alignment, and (c) meet the band alignment
criteria. Following this, we infer that these three heterostructures
are promising candidates for efficient photocatalysis.

\begin{table}[H]

\caption{\protect\label{tab:hetero-bandedges}Calculated values of band edges,
$E_{CBM}$ and $E_{VBM}$, of the considered TMD heterojunctions at
the HSE06 and G$_{0}$W$_{0}$ level. The values are calculated with
respect to vacuum level.}

\centering{}%
\begin{tabular}{cccccc}
\toprule 
\multirow{2}{*}{S.No.} & \multirow{2}{*}{Heterostructures} & \multicolumn{2}{c}{$E_{VBM}$ (eV)} & \multicolumn{2}{c}{$E_{CBM}$ (eV)}\tabularnewline
\cmidrule{3-6}
 &  & HSE06 & G$_{0}$W$_{0}$ & HSE06 & G$_{0}$W$_{0}$\tabularnewline
\midrule
\midrule 
1 & WS\textsubscript{2}/MoS\textsubscript{2} & -5.643 & -5.813 & -3.890 & -3.662\tabularnewline
2 & MoSe\textsubscript{2}/MoS\textsubscript{2} & -6.615 & -5.727 & -5.040 & -3.834\tabularnewline
3 & WSe\textsubscript{2}/MoS\textsubscript{2} & -5.475 & -5.775 & -4.001 & -3.698\tabularnewline
4 & WSe\textsubscript{2}/MoSe\textsubscript{2} & -5.154 & -5.384 & -3.718 & -3.444\tabularnewline
5 & WSe\textsubscript{2}/WS\textsubscript{2} & -5.445 & -5.614 & -3.972 & -3.459\tabularnewline
6 & MoSe\textsubscript{2}/WS\textsubscript{2} & -5.424 & -5.556 & -3.825 & -3.619\tabularnewline
\bottomrule
\end{tabular}
\end{table}

\section{\protect\label{sec:Conclusion}Conclusion}

To summarize, we have proposed 2D transition-metal dichalcogenides
based vertical bilayer heterostructures for solar cell devices using
first-principles DFT and many body perturbation based G$_{0}$W$_{0}$--
BSE calculations. Using four TMD monolayers (MoS\textsubscript{2},
WS\textsubscript{2}, MoSe\textsubscript{2}, and WSe\textsubscript{2})
which satisfy the lattice mismatch condition, six such heterostructures
can be created. Among the six possible heterostructures, five namely,
WS\textsubscript{2}/MoS\textsubscript{2}, MoSe\textsubscript{2}/MoS\textsubscript{2},
WSe\textsubscript{2}/MoS\textsubscript{2}, WSe\textsubscript{2}/MoSe\textsubscript{2},
and WSe\textsubscript{2}/WS\textsubscript{2}, meet the criterion
of type II band alignment according to both the HSE06 functional and
G$_{0}$W$_{0}$ approximation, whereas, MoSe\textsubscript{2}/WS\textsubscript{2}
shows type II band alignment at the HSE06 level only. We have examined
the structural stability of the considered heterostructures, after
which a detailed systematic investigation of their electronic structures
is performed using both the DFT and G$_{0}$W$_{0}$ approximation.
As compared to the isolated monolayers, the considered heterostructures
exhibit reduced electronic gaps, leading to wider coverage of solar
spectrum. The spatial separation of photogenerated charge carriers,
which is confirmed by the orbital projected PBE band structures, will
lower the electron-hole recombination rate. The calculated values
of the electron and hole effective masses are small and nearly equal,
which will facilitate their flow with same mobility in the opposite
directions, further lowering the recombination rates. The built-in
electric fields, which arises due to the horizontal mirror asymmetry
in the considered heterostructures, are also computed, and found to
be quite significant. It is noteworthy that these intrinsic electric
fields will also lead to charge separation, and, thus, lower recombination
rates. The optical responses are obtained by solving the BSE, and
the PCEs are also computed at both the HSE06 and G$_{0}$W$_{0}$
level. The PCE is not evaluated for MoSe\textsubscript{2}/WS\textsubscript{2}
at the G$_{0}$W$_{0}$ level because it does not manifest type II
band alignment. Our study demonstrates that by carefully designing
Type II heterostructures using TMD monolayers, it is possible to tailor
their properties so that they will exhibit low recombination losses
and high PCEs (near 20\% at the HSE06 level) leading to superior solar
cells. 

Afterwards, the six TMD heterostructures under consideration are evaluated
for their potential to function as photocatalysts for hydrogen evolution
via water splitting, and WS\textsubscript{2}/MoS\textsubscript{2},
MoSe\textsubscript{2}/MoS\textsubscript{2},
and WSe\textsubscript{2}/MoS\textsubscript{2}
heterostructures are found to be quite promising. 

In this paper we have made a number of predictions regarding the photovoltaic
and photocatalytic abilities of the six TMD heterostructures, and
we hope they will be tested in future experiments.

Moreover, in this work, we considered no relative rotation between
the monolayers constituting the heterostructures. However, given the
current interest in Moire physics induced by a relative twist angle
between the layers, in future we intend to undertake a systematic
study of its influence on various properties of such TMD heterojunctions. 

\section*{acknowledgments}

One of the authors, K.D. acknowledges financial assistance from the
Prime Minister Research Fellowship (PMRF ID-1302054), MHRD, India,
and Space Time computational facility of Indian Institute of Technology
Bombay. R.Y. acknowledges the support through the Institute Post-Doctoral
Fellowship (IPDF) of Indian Institute of Technology Bombay.

\bibliographystyle{unsrt}
\bibliography{Ref}

\end{document}


\title{Supplemental Material for 2D Transition-metal dichalcogenides based
bilayer heterojunctions for efficient solar cells and photocatalytic
applications }
\author{Khushboo Dange}
\email{khushboodange@gmail.com}

\affiliation{Department of Physics, Indian Institute of Technology Bombay, Powai,
Mumbai 400076, India}
\author{Rachana Yogi}
\email{yogirachana04@gmail.com}

\affiliation{Department of Physics, Indian Institute of Technology Bombay, Powai,
Mumbai 400076, India}
\author{Alok Shukla}
\email{shukla@iitb.ac.in}

\affiliation{Department of Physics, Indian Institute of Technology Bombay, Powai,
Mumbai 400076, India}
\maketitle

\subsection{TMD Monolayers}

\begin{figure}[H]
\begin{centering}
\includegraphics[scale=0.3]{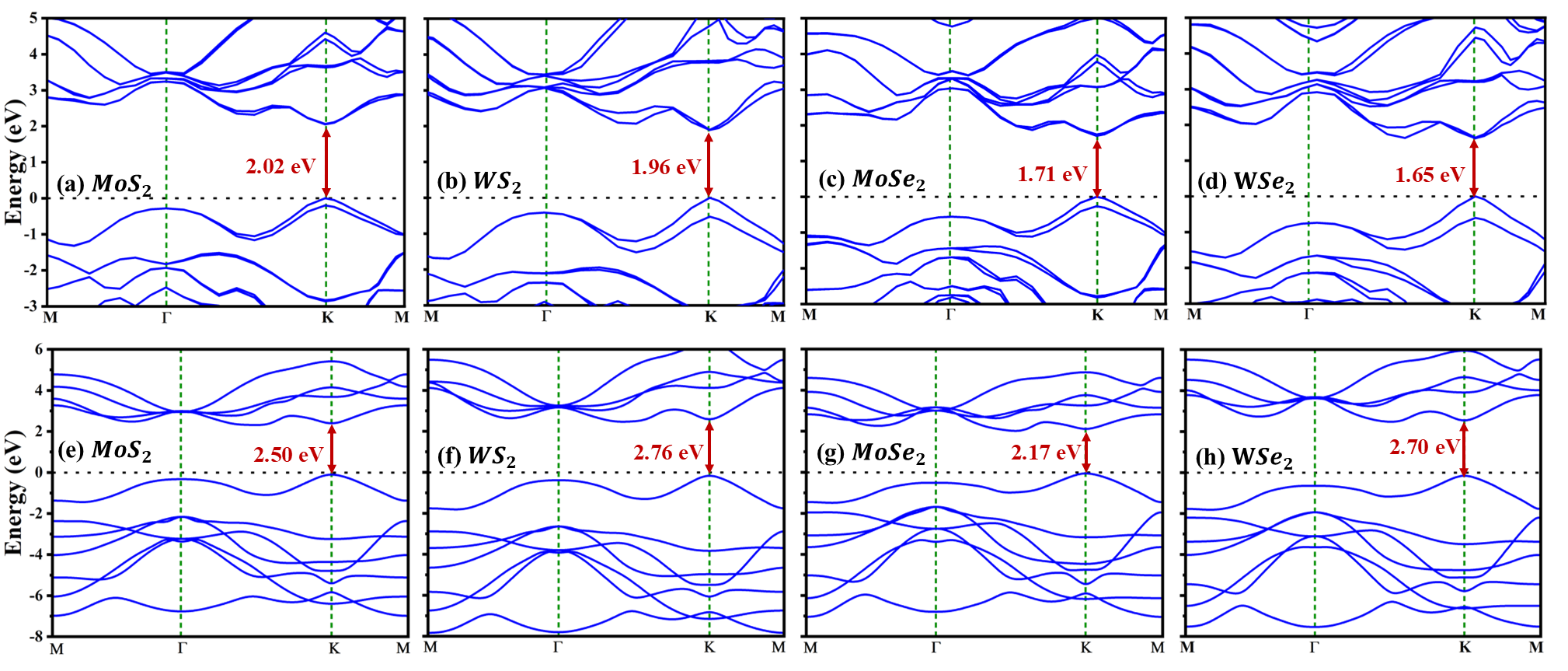}
\par\end{centering}
\caption{Band structures of the considered TMD monolayers, calculated using
(a)-(d) HSE06 functional and (e)-(f) G$_{0}$W$_{0}$ approximation.}

\end{figure}

\begin{table}[H]
\caption{Reported carrier mobilities ($\mu$) of the 2D TMDs under consideration}

\begin{centering}
\begin{tabular}{ccc}
\toprule 
2D TMDs & $\mu$ (10$^{3}$cm$^{3}$V$^{-1}$S$^{-1}$) & References\tabularnewline
\midrule
\midrule 
MoS\textsubscript{2} & 0.410 & \citep{li2019}\tabularnewline
MoSe\textsubscript{2} & 0.240 & \citep{Zhang2014}\tabularnewline
WS\textsubscript{2} & 1.102 & \citep{li2019}\tabularnewline
WSe\textsubscript{2} & 0.705 & \citep{li2019}\tabularnewline
\bottomrule
\end{tabular}
\par\end{centering}
\end{table}

\begin{figure}[H]
\begin{centering}
\includegraphics[scale=0.6]{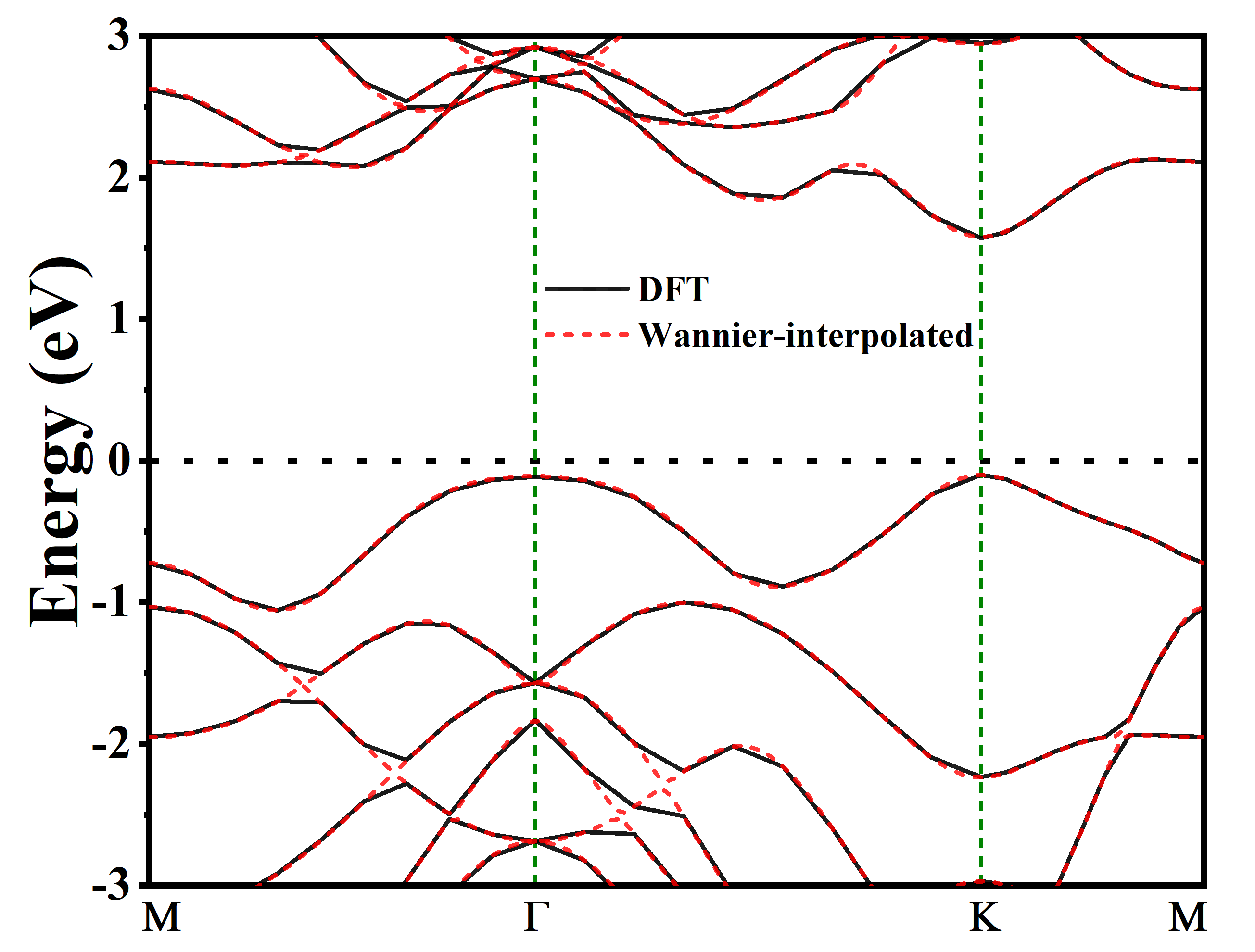}
\par\end{centering}
\caption{GGA-PBE (without SOC) band structure of the MoS\protect\textsubscript{2}
monolayer.}
\end{figure}

\begin{figure}
\begin{centering}
\includegraphics[scale=0.35]{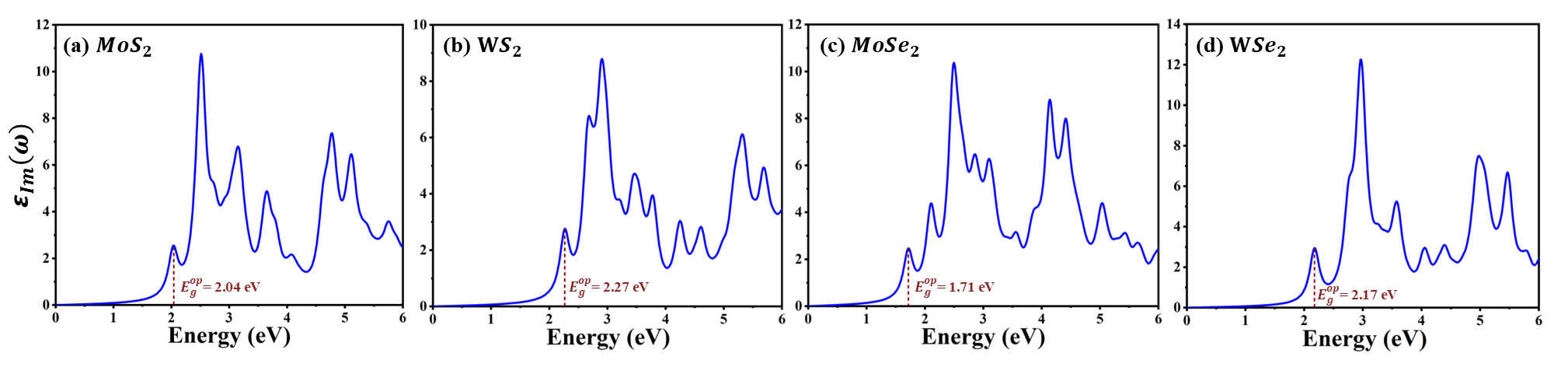}
\par\end{centering}
\caption{Imaginary part of dielectric function, $\epsilon_{Im}(\omega)$ as
a function of energy of the incident photon for the considered monolayers
using BSE.}
\end{figure}

\begin{table}

\caption{Calculated effective masses of the electrons ($m_{e}^{*}$) and holes
($m_{h}^{*}$) in terms of free electron mass ($m_{0}$) for the considered
TMD monolayers. }

\centering{}%
\begin{tabular}{ccccc}
\toprule 
\multirow{2}{*}{Monolayers} & \multicolumn{2}{c}{$m_{e}^{*}$($m_{0}$)} & \multicolumn{2}{c}{$m_{h}^{*}(m_{0})$}\tabularnewline
\cmidrule{2-5}
 & (K$\rightarrow\Gamma$) & (K$\rightarrow$M) & (K$\rightarrow\Gamma$) & (K$\rightarrow$M)\tabularnewline
\midrule 
MoS$_{2}$ & 0.48 & 0.54 & 0.56 & 0.70\tabularnewline
WS$_{2}$ & 0.32 & 0.34 & 0.42 & 0.49\tabularnewline
MoSe$_{2}$ & 0.55 & 0.63 & 0.63 & 0.79\tabularnewline
WSe$_{2}$ & 0.35 & 0.39 & 0.45 & 0.53\tabularnewline
\bottomrule
\end{tabular}
\end{table}

\subsection{TMD heterostructures}

\begin{figure}[H]
\begin{centering}
\includegraphics[scale=0.35]{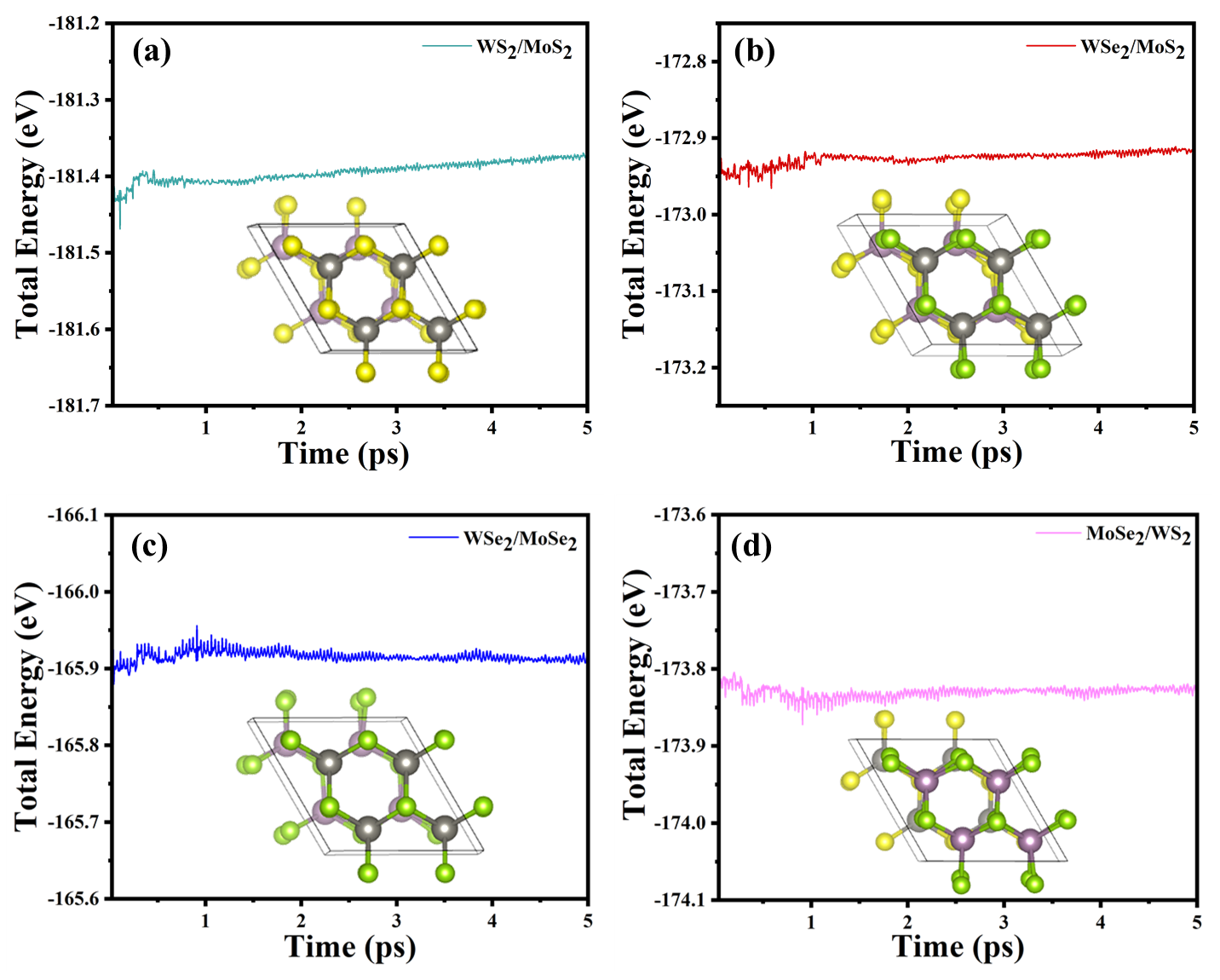}
\par\end{centering}
\caption{Total energy as a function of time steps for (a) WS\protect\textsubscript{2}/MoS\protect\textsubscript{2},
(b) WSe\protect\textsubscript{2}/MoS\protect\textsubscript{2}, (c)
WSe\protect\textsubscript{2}/MoSe\protect\textsubscript{2}, and
(d) MoSe\protect\textsubscript{2}/WS\protect\textsubscript{2} systems,
obtained from the AIMD simulations at 500K. The final structures obtained
at the end of simulations are also shown in the inset. Here, purple,
grey, yellow, and green spheres represent the Mo, W, S, and Se atoms,
respectively.}

\end{figure}

\begin{figure}[H]
\begin{centering}
\includegraphics[scale=0.3]{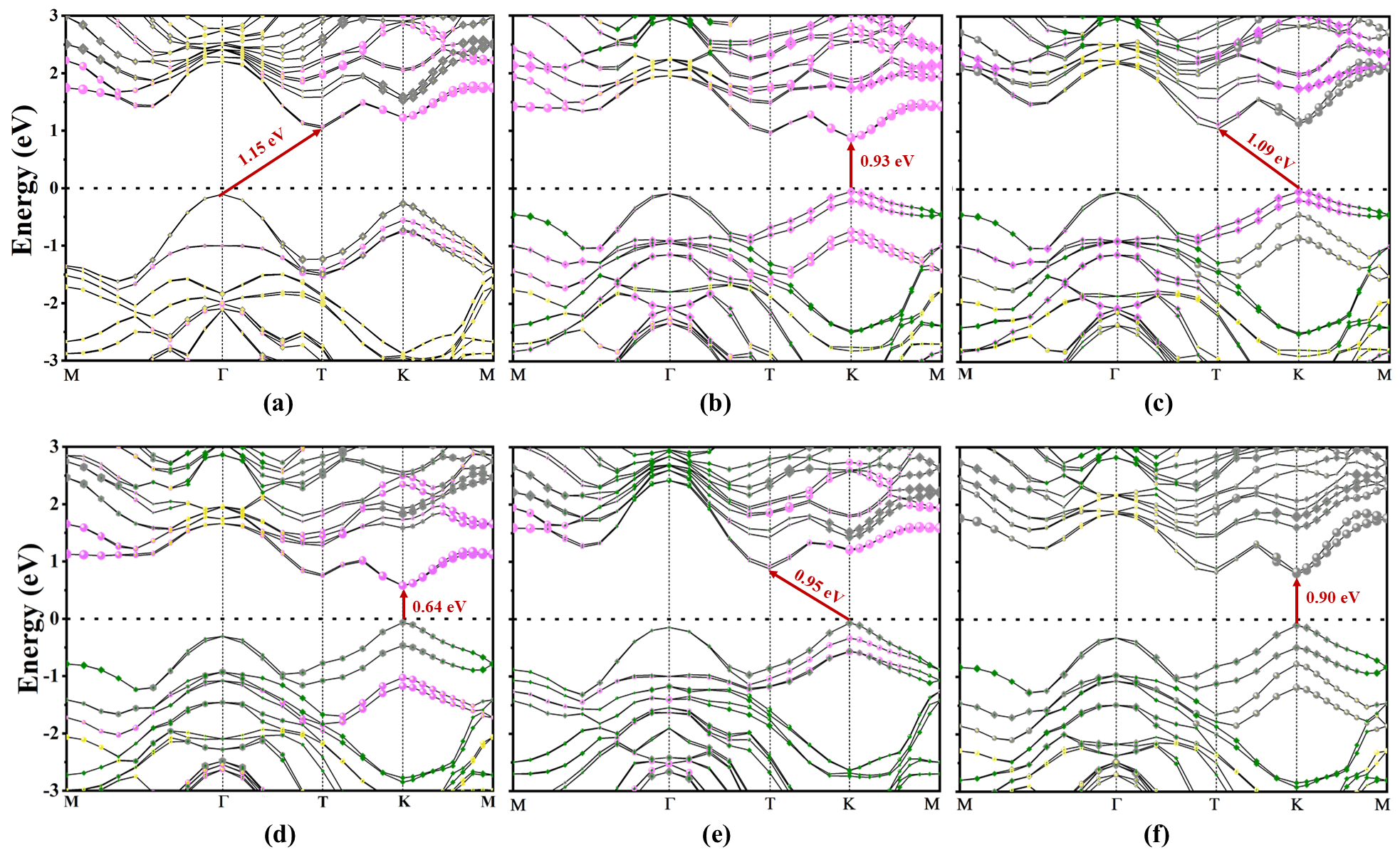}
\par\end{centering}
\caption{Projected band structures of (a) WS\protect\textsubscript{2}/MoS\protect\textsubscript{2},
(b) MoSe\protect\textsubscript{2}/MoS\protect\textsubscript{2},
(c) MoSe\protect\textsubscript{2}/WS\protect\textsubscript{2}, (d)
WSe\protect\textsubscript{2}/MoS\protect\textsubscript{2}, (e) WSe\protect\textsubscript{2}/MoSe\protect\textsubscript{2}
and (f) WSe\protect\textsubscript{2}/WS\protect\textsubscript{2}
heterojunctions using PBE functional. Pink, grey, yellow, and green
spheres (diamonds) denote the Mo, W, S, and Se atoms of the acceptor
(donor) semiconductors.}
\end{figure}

\begin{table}
\caption{$E_{g}^{QP}$ values using 200 unoccupied bands in G$_{0}$W$_{0}$
calculations}

\centering{}%
\begin{tabular}{ccc}
\toprule 
S.No. & Heterostructures & $E_{g}^{QP}$(eV)\tabularnewline
\midrule 
1 & WS\textsubscript{2}/MoS\textsubscript{2} & 2.13\tabularnewline
2 & MoSe\textsubscript{2}/MoS\textsubscript{2} & 1.92\tabularnewline
3 & MoSe\textsubscript{2}/WS\textsubscript{2} & 1.93\tabularnewline
4 & WSe\textsubscript{2}/MoS\textsubscript{2} & 2.07\tabularnewline
5 & WSe\textsubscript{2}/MoSe\textsubscript{2} & 1.98\tabularnewline
6 & WSe\textsubscript{2}/WS\textsubscript{2} & 2.14\tabularnewline
\bottomrule
\end{tabular}
\end{table}

\begin{figure}
\begin{centering}
\includegraphics[scale=0.6]{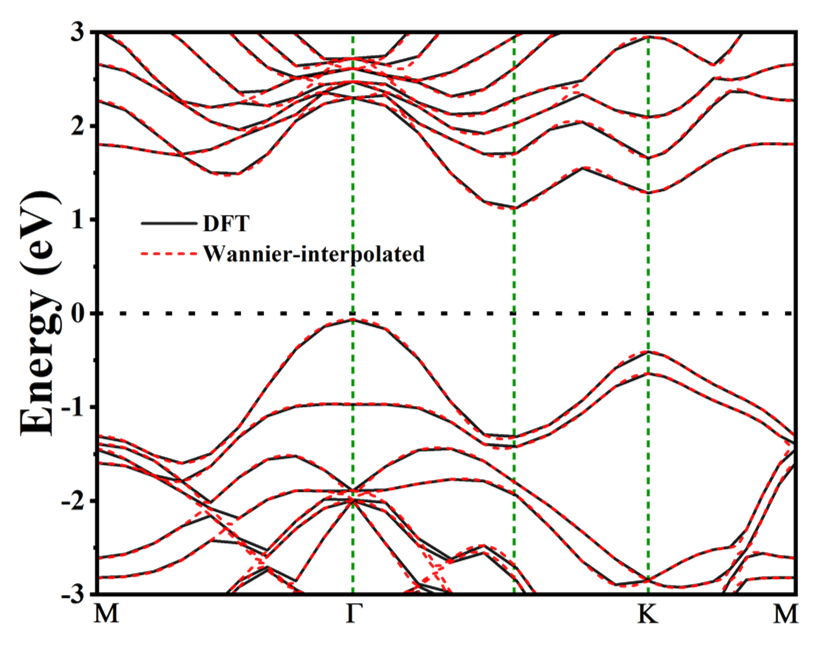}
\par\end{centering}
\caption{GGA-PBE band structure (without SOC) of the WS\protect\textsubscript{2}/MoS\protect\textsubscript{2}
heterostructure.}

\end{figure}
\begin{figure}[H]
\begin{centering}
\includegraphics[scale=0.5]{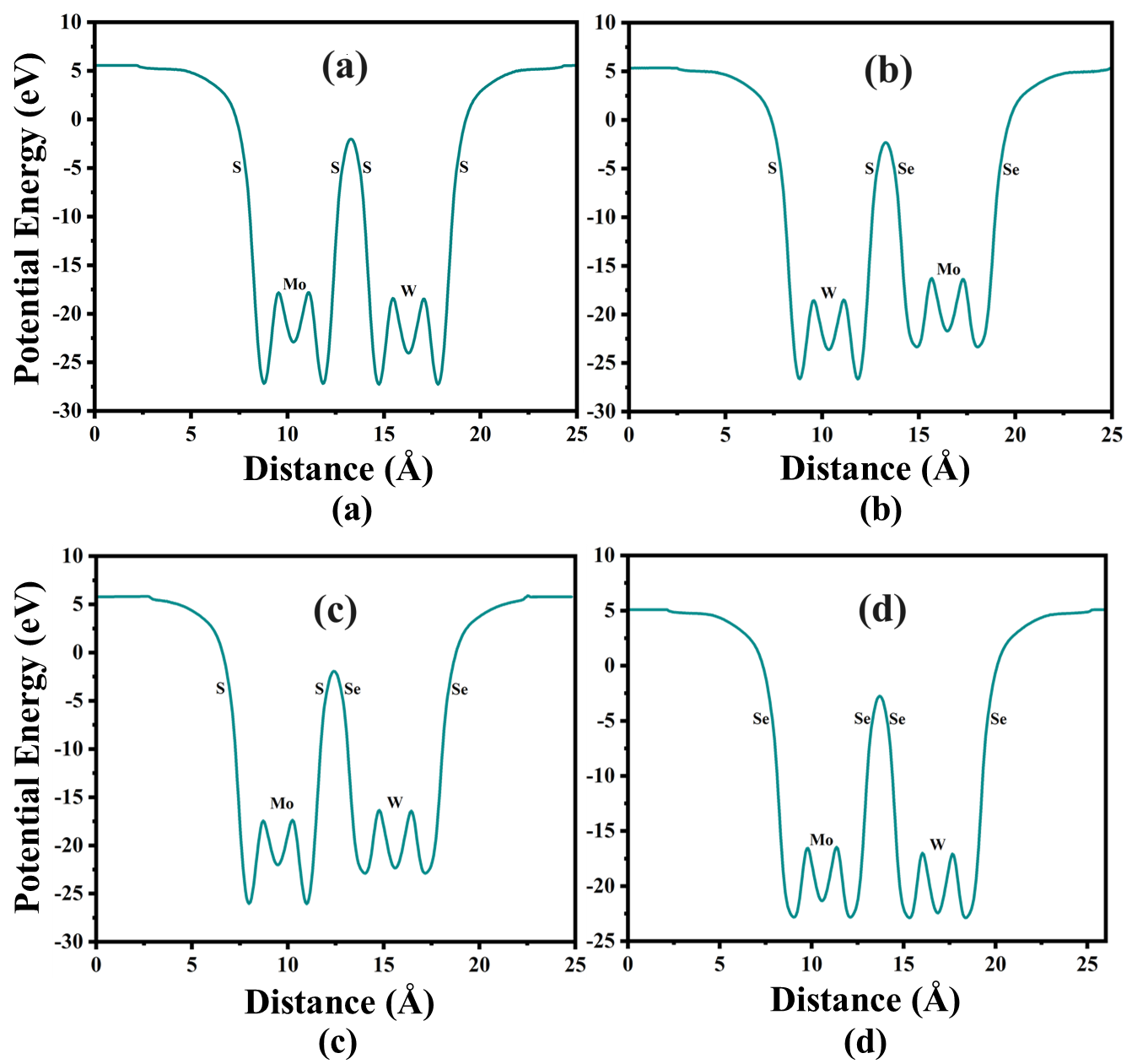}
\par\end{centering}
\caption{Electrostatic potential energy as a function of distance along the
$z$-direction for (a) WS\protect\textsubscript{\textcolor{black}{2}}/MoS\protect\textsubscript{\textcolor{black}{2}},
(b) MoSe\protect\textsubscript{\textcolor{black}{2}}/WS\protect\textsubscript{\textcolor{black}{2}},
(c) WSe\protect\textsubscript{\textcolor{black}{2}}/MoS\protect\textsubscript{\textcolor{black}{2}},
and (d) WSe\protect\textsubscript{\textcolor{black}{2}}/MoSe\protect\textsubscript{\textcolor{black}{2}}
heterojunctions.}
\end{figure}

\begin{table}[H]
\centering{}\caption{Solar cell parameters, donor band gap ($E_{g}^{d}$ ), conduction
band offset ($\Delta E_{c}$), open circuit voltage ($V_{oc}$), and
current density ($J_{sc}$), of the heterostructures under consideration
calculated at DFT level using HSE06 functional.}
\begin{tabular}{cccccc}
\toprule 
S.No. & Heterostructures & $E_{g}^{d}$ (eV) & $\Delta E_{c}$(eV) & $V_{oc}$(V) & $J_{sc}$(mA/cm\textsuperscript{2})\tabularnewline
\midrule
\midrule 
1 & WS\textsubscript{2}/MoS\textsubscript{2} & 1.96 & 0.37 & 1.28 & 15.48\tabularnewline
2 & MoSe\textsubscript{2}/MoS\textsubscript{2} & 1.71 & 0.12 & 1.29 & 22.16\tabularnewline
3 & MoSe\textsubscript{2}/WS\textsubscript{2} & 1.71 & 0.17 & 1.23 & 22.16\tabularnewline
4 & WSe\textsubscript{2}/MoS\textsubscript{2} & 1.65 & 0.43 & 0.92 & 23.98\tabularnewline
5 & WSe\textsubscript{2}/MoSe\textsubscript{2} & 1.65 & 0.31 & 1.04 & 23.98\tabularnewline
6 & WSe\textsubscript{2}/WS\textsubscript{2} & 1.65 & 0.11 & 1.24 & 23.98\tabularnewline
\bottomrule
\end{tabular}
\end{table}

\begin{table}[H]
\centering{}\caption{Solar cell parameters, donor band gap ($E_{g}^{d}$ ), conduction
band offset ($\Delta E_{c}$), open circuit voltage ($V_{oc}$), and
current density ($J_{sc}$), of the heterostructures under consideration
calculated at G$_{0}$W$_{0}$ level.}
\begin{tabular}{cccccc}
\toprule 
S.No. & Heterostructures & $E_{g}^{d}$ (eV) & $\Delta E_{c}$(eV) & $V_{oc}$(V) & $J_{sc}$(mA/cm\textsuperscript{2})\tabularnewline
\midrule
\midrule 
1 & WS\textsubscript{2}/MoS\textsubscript{2} & 2.27 & 0.43 & 1.54 & 9.39\tabularnewline
2 & MoSe\textsubscript{2}/MoS\textsubscript{2} & 1.71 & 0.38 & 1.03 & 22.16\tabularnewline
3 & WSe\textsubscript{2}/MoS\textsubscript{2} & 2.17 & 0.96 & 0.91 & 11.10\tabularnewline
4 & WSe\textsubscript{2}/MoSe\textsubscript{2} & 2.17 & 0.58 & 1.29 & 11.10\tabularnewline
5 & WSe\textsubscript{2}/WS\textsubscript{2} & 2.17 & 0.53 & 1.34 & 11.10\tabularnewline
\bottomrule
\end{tabular}
\end{table}

\begin{table}[H]
\centering{}\caption{Solar cell parameters, donor band gap ($E_{g}^{d}$ ), conduction
band offset ($\Delta E_{c}$), open circuit voltage ($V_{oc}$), and
current density ($J_{sc}$), of the heterostructures under consideration
calculated at G$_{0}$W$_{0}$+ SOC level.}
\begin{tabular}{cccccc}
\toprule 
S.No. & Heterostructures & $E_{g}^{d}$ (eV) & $\Delta E_{c}$(eV) & $V_{oc}$(V) & $J_{sc}$(mA/cm\textsuperscript{2})\tabularnewline
\midrule
\midrule 
1 & WS\textsubscript{2}/MoS\textsubscript{2} & 2.01 & 0.39 & 1.32 & 14.34\tabularnewline
2 & MoSe\textsubscript{2}/MoS\textsubscript{2} & 1.60 & 0.38 & 0.92 & 25.44\tabularnewline
3 & WSe\textsubscript{2}/MoS\textsubscript{2} & 1.87 & 0.92 & 0.65 & 17.68\tabularnewline
4 & WSe\textsubscript{2}/MoSe\textsubscript{2} & 1.87 & 0.54 & 1.03 & 17.68\tabularnewline
5 & WSe\textsubscript{2}/WS\textsubscript{2} & 1.87 & 0.53 & 1.04 & 17.68\tabularnewline
\bottomrule
\end{tabular}
\end{table}

\bibliographystyle{plain}
\bibliography{Ref_SI}